\documentclass[11pt]{article}
\usepackage[dvips]{graphicx}
\begin{document}
\tolerance=10000
\hyphenpenalty=2000
\hyphenation{visco-elastic visco-elasticity}
\setcounter{page}{1}

\def\pni{\par \noindent}
\def\vsh{\smallskip \noindent}
\def\vs{\medskip \noindent}
\def\vvs{\bigskip\noindent}
\def\vvvs{\bigskip\medskip} 
\def\vsp{\vsh\pni}
\def\vsn{\vsh\pni}
\def\cen{\centerline}
\centerline{{FRACALMO PRE-PRINT} \ {\bf www.fracalmo.org}}
\vspace{0.1truecm}
 \centerline{\bf Chemical Physics, Vol. 284 No 1/2 (2002) pp. 521-541.}
\centerline{Special Issue "Stange Kinetics"}
\centerline{Guest Editors: R. Hilfer, R. Metzler, A. Blumen, J. Klafter}
\vspace{0.1 truecm} \hrule \vspace{0.1truecm}
\font\title=cmbx12 scaled\magstep2
\font\bfs=cmbx12 scaled\magstep1
\font\little=cmr10

\begin{center}
{\title Discrete random walk models }

\vs
{\title for space-time fractional diffusion}

\vvs
{Rudolf GORENFLO}$^{(1)}$, {Francesco MAINARDI}$^{(2)}$ 

\vs
{Daniele MORETTI}$^{(3)}$, {Gianni PAGNINI}$^{(4)}$, {Paolo PARADISI}$^{(5)}$ 

\vs
$\null^{(1)}$
 {\little First Mathematical Institute,
 Free   University of Berlin,} \\
{\little  Arnimallee  3, D-14195 Berlin, Germany} 
\\ [0.20 truecm]
$\null^{(2)}$
 {\little Dipartimento di Fisica, Universit\`a di Bologna and INFN,} \\
{\little Via Irnerio 46, I-40126 Bologna, Italy} 
\\ [0.20 truecm]
$\null^{(3)}$
{\little Studio PLEIADI, Via Cairoli 35, I-47023 Cesena, Italy}
\\ [0.20 truecm]
$\null^{(4)}$
{\little ENEA: Italian Agency for  New Technologies,
  Energy and the Environment} 
\\ {\little Via Martiri di Monte Sole 4,I-40129 Bologna, Italy}
   \\ [0.20 truecm]
$\null^{(5)}$   
{\little ISAC: Istituto per le Scienze dell'Atmosfera e del Clima
    del CNR,} \\ 
{\little Strada Provinciale Lecce-Monteroni Km 1.200, I-73100 Lecce, Italy}
\\ [0.20truecm]
\end{center}

\cen{\bf Abstract} 

\vskip 0.1truecm
\noindent
A  physical-mathematical 
approach to anomalous diffusion
may be	based on generalized diffusion equations
(containing derivatives of fractional order in space or/and time)
and related random walk models.
 By   space-time fractional diffusion
equation we mean an evolution equation obtained from the standard linear
diffusion equation by replacing the second-order space derivative
with a Riesz-Feller derivative of order $\alpha \in (0,2]$
and skewness $\theta$ ($|\theta|\le\hbox{min}\,\{\alpha ,2-\alpha \}$),
and the first-order time derivative with a Caputo derivative
of order $\beta \in (0,1]\,.$
Such evolution equation implies for the flux a fractional Fick's law
which accounts for spatial and temporal non-locality.
The fundamental solution (for the {Cauchy} problem)
of the fractional diffusion equation  can be interpreted as
a probability density  evolving in time
of a peculiar {self-similar}  stochastic process
that we view as a generalized diffusion process.
By adopting  appropriate
finite-difference schemes  of solution, we generate
models of random walk discrete in space and time
suitable for simulating
random variables whose spatial probability density evolves
in time  according to  this fractional diffusion equation.
\newpage
\vs

\vs


\noindent
\newcommand{\NN}{\bf{N}}
\newcommand{\ZZ}{\bf{Z}}
\newcommand{\CC}{\bf{C}}
\newcommand{\RR}{\bf{R}}
 \newcommand{\intl}{\int\limits}
\newcommand{\suml}{\sum\limits}

\def\eg{{\it e.g.}\ } \def\ie{{\it i.e.}\ }
\def\sg{\hbox{sign}\,}
\def\sgn{\hbox{sign}\,}
\def\sign{\hbox{sign}\,}
\def\e{\hbox{e}}
\def\exp{\hbox{exp}}
\def\ds{\displaystyle}
\def\dis{\displaystyle}
\def\q{\quad}	 \def\qq{\qquad}
\def\lan{\langle}\def\ran{\rangle}
\def\l{\left} \def\r{\right}
\def\lra{\Longleftrightarrow}
\def\arg{\hbox{\rm arg}}
\def\d{\partial}
 \def\dr{\partial r}  \def\dt{\partial t}
\def\dx{\partial x}   \def\dy{\partial y}  \def\dz{\partial z}
\def\rec#1{{1\over{#1}}}
\def\log{\hbox{\rm log}\,}
\def\erf{\hbox{\rm erf}\,}     \def\erfc{\hbox{\rm erfc}\,}
\def\zr{z^{-1}}    
\def\G{{G_{\alpha,\beta}^\theta}}
\def\K{K_{\alpha,\beta}^\theta}
\def\Gxt{\G (x,t)}
\def\Gkt{{\widehat{\G}}  (\kappa,t)}
\def\Gxs{{\widetilde{\G}}  (x,s)}
\def\Gks{{\widehat{\widetilde {\G}}} (\kappa,s)}
\def\FT{{\cal F}\,} 
\def\LT{{\cal L}\,}  
\def\L{{\cal L}} 
\def\F{{\cal F}} 
\def\M{{\cal M}}  
\def\I{{\cal I}}  
\def\ra{\item{a)\ }} \def\rb{\item{b)\ }}   \def\rc{\item{c)\ }}
\def\alphak{{\alpha \choose k}}
\def\alphazero{{\alpha \choose 0}}
\def\alphaone{{\alpha \choose 1}}
\def\alphatwo{{\alpha \choose 2}}
\def\alphakk{{\alpha \choose k+1}}


\section{Introduction}

It is well known that the fundamental solution (or {\it Green function})
for the {Cauchy} problem of the   linear diffusion equation
can be interpreted as a Gaussian (normal)  probability density function ($pdf$)
in space, evolving in time.
All the moments of this $pdf$ are finite;
in particular, its variance
is proportional to the first power of time, a noteworthy property
of the {\it standard diffusion} that can
be understood by  means of an unbiased random walk model for
the {\it Brownian motion}.
\vsp
In recent years a number of master equations have been proposed
for random walk models that turn out to be
beyond the classical Brownian motion,
see \eg Klafter {\it et al}. \cite{Klafter PhysToday96}. 
In particular, evolution equations containing fractional derivatives
have gained revived interest in that they are expected to provide
suitable mathematical models  for describing phenomena of
anomalous diffusion  and transport dynamics in complex systems, see \eg
\cite{Barkai PRE01},
\cite{ChechkinGonchar JETP00},
\cite{GionaRoman PhysA92},
\cite{HenryWearne 00}
\cite{Hilfer 00a,HilferAnton 95},
\cite{Mainardi WASCOM93,Mainardi CHAOS96,Mainardi CISM97},
\cite{Metzler PhysA94},
\cite{Paradisi PhysA01,Paradisi PhysChem01},
\cite{SaichevZaslavsky 97},
\cite{SchneiderWyss 89},
\cite{Uchaikin 00,UchaikinZolotarev 99},
\cite{WestGrigoliniMetzlerTheo PRE97,WestSeshadri PhysA82}.
  For a recent 
review we refer  to Metzler and Klafter
\cite{MetzlerKlafter PhysRep00}
where  other references  are found. 
\vsp
Here we   intend to
present our original approach to the topic
that, being not considered in	\cite{MetzlerKlafter PhysRep00},
could offer some novel and inspiring inspections
to the phenomenon of anomalous diffusion
which is of great interest in chemical physics.
In this paper we  complement and revisit
some of our previous  results found \eg in 
\cite{GorDFMai PhysA99}, \cite{GorMai FCAA98}, \cite{GorMai JVC02}. 
We first show  that
our  proposed fractional diffusion equations
can be derived from  generalized
Fick's laws which account for spatial and/or  temporal non-locality.
Then we pay attention to the fact that
the fundamental solutions (or {\it Green functions}) of our
diffusion equations	
provide spatial  probability densities evolving in
time,	related to  {\it self-similar} stochastic processes,
that we view as generalized (or {\it fractional})
diffusion processes to be properly understood through
suitable random walk models.
More precisely, we  replace
the second-order space derivative or/and the first-order time derivative
by   a suitable {\it integro-differential} operator, which can be
interpreted as a space or time derivative of fractional order
$\alpha \in (0,2]$ 
or $\beta \in (0,1]\,,$ 
respectively
\footnote{We remind that the term "fractional" is a misnomer
since the order can be a real number and thus is not restricted to  be
rational. The term is kept only for historical reasons,
see \eg \cite{GorMai CISM97}.
Our fractional derivatives are required to
coincide with the standard derivatives of integer order as soon as
$\alpha=2$ (not as $\alpha =1$!)   and $\beta =1\,. $}.
The space fractional derivative
is required to depend also on a real parameter $\theta$
(the {\it skewness})  subjected to the
restriction $|\theta|\le\hbox{min}\,\{\alpha ,2-\alpha \}\,. $
Correspondingly, the   generalized equation will be referred to as
the {\it strictly space fractional diffusion} equation of order $\alpha $
and skewness $\theta$ if  $\alpha\in (0,2)$  
and $\beta =1\,, $
or the	  {\it	strictly time fractional diffusion} equation of order
$\beta \, $ if $\alpha =2$  and $\beta \in (0,1).$  
In general,   allowing 
$\alpha \in (0,2)$ and $\beta \in (0,1)\,,$
we have the {\it strictly space-time fractional diffusion} equation of
order $\alpha,\beta$  and
skewness $\theta\,. $
Of course, in the case $\{\alpha=2,\; \; \beta=1 \}$
we recover the {\it standard diffusion} which leads to the Gaussian
probability density and to the classical {\it Brownian motion}.
\vsp
For the {\it strictly space fractional diffusion} 
of order $\alpha$   ($\{0<\alpha < 2, \, \beta =1\}$)
we generate the class of (non-Gaussian) L\'evy stable densities
of index $\alpha$   and  skewness $\theta$
($|\theta| \le\hbox{min}\,\{\alpha ,2-\alpha \}$),
according to the Feller parameterization.
As known, these  densities  exhibit
fat tails with an algebraic decay $\propto |x|^{-(\alpha +1)}\,.$
We thus obtain a special class of {\it Markovian}
stochastic processes, called
{\it stable L\'evy motions}, which exhibit infinite variance
associated to the possibility of arbitrarily large jumps
({\it L\'evy flights}).
\vsp
For the {\it strictly time fractional diffusion} 
of order $\beta $  ($\{\alpha  =2,\, 0<\beta  < 1\}$)
we generate   a class of symmetric densities
whose  moments of order $2n$ are proportional to
the  $n\, \beta$ power	 of time.
We thus obtain a  class of {\it non-Markovian} stochastic processes
(they possesses a memory!)
which exhibit a variance consistent with {\it slow anomalous diffusion}.
\vsp
For the {\it strictly space-time fractional diffusion} 
of 
(composite) order $\alpha \in (0,2),$ $\,\beta\in (0,1) $
we generate   a class of  densities
(symmetric or not symmetric according to $\theta =0$ or $\theta \ne 0$)
which exhibit  fat tails
with an algebraic decay $\propto |x|^{-(\alpha +1)}\,.$
Thus they belong to the domain of attraction of the L\'evy stable densities
of index $\alpha $ and	can be called
{\it fractional stable densities}.
The related stochastic processes,
by possessing the  characteristics of the previous two classes,
are {\it non-Markovian} and exhibit infinite variance;
however, the possibility of arbitrary large jumps is
contrasted by  memory effects.
Furthermore we mention the cases $\alpha =\beta $ for which
it is possible to derive the Green function in closed
analytical form: we refer to these cases as to {\it neutral
diffusion}.
\newpage
\vsp
We shall prove that in any case the corresponding Green function
can be interpreted as a spatial probability density
evolving in time
with a	self-similarity property having scaling
exponent $\nu  = \beta /\alpha \,. $
This allows us to limit ourselves to consider
the expression of the Green function at a fixed time, say $t=1\,,$
namely to the so-called {\it reduced Green function}.
To approximate the time evolution  of all the
above  densities
we propose  finite difference schemes, discrete
in space and time, for the fractional derivatives.
 By taking care in constructing these schemes,
namely by requiring them to  be conservative and non-negativity
preserving,
they can be interpreted
 as discrete random walk models for simulating particle paths
 by the Monte Carlo technique.
 By properly scaled transition to vanishing space and time steps,
 these models can be shown to converge to the corresponding continuous
 processes\footnote{
This was shown by Gorenflo and Mainardi for the  space fractional
diffusion in \cite{GorMai FCAA98,GorMai ZAA99,GorMai CHEMNITZ01}.
For the general case it will be shown in a next paper.}.
\vsp
The paper is divided as follows.
In Section 2  we first present our space and time
fractional diffusion
equations providing the definitions of the space and time fractional
derivatives based on their Fourier and Laplace
representations, respectively.
Then we show how to derive them from  generalized
Fick's laws.
Section 3 is devoted to 
 the Green functions, pointing out their
similarity properties. We provide  the representations
of the corresponding
{\it reduced Green functions} in terms of Mellin-Barnes integrals
which	allow us to obtain their computational expressions.
In Section 4, we first discuss
the discrete random walk approach to the Brownian motion,
which is based on the well-known discretization
of the second order space derivative and the first-order time
derivative entering the standard diffusion equation.
Then,  by properly discretizing the space-fractional derivative
we generalize the above approach to    the more general
Markovian case
of {\it strictly space fractional diffusion}.
Section 5 is devoted to the extension of the above
approach to the non-Markovian cases  of {\it strictly time fractional
diffusion}  and {\it strictly space-time fractional diffusion} equations.
Section 6 is devoted to the numerical results
of our random walks produced
in some case-studies and to  the concluding discussions.
For possible convenience of the reader we have reserved
Appendix A and Appendix B for treating with some detail the Riesz-Feller
and Caputo fractional derivatives, respectively.

\section{The space-time fractional diffusion equation}

By  replacing in the standard diffusion equation
$$ {\d\over \dt} u(x,t) =  {\d^2\over \dx^2}\,
  u(x,t)\,,
   \q -\infty< x <+\infty\,, \q t \ge 0\,,
   \eqno(2.1)$$
where $u=u(x,t)$ is the (real) field variable,
the second-order space derivative and the first-order time derivative
by    suitable {\it integro-differential} operators, which can be
interpreted as a space and time derivative of fractional order
we  obtain a   generalized diffusion equation which may be
referred
to as the {\it space-time fractional} diffusion equation.
We write this equation as
$$
_tD_*^\beta \, u(x,t) \,       = \, _xD_\theta^\alpha \,u(x,t) \,,
\q -\infty< x <+\infty\,, \q t \ge 0\,,
\eqno(2.2) $$
where   $\alpha \,,\,\theta\,,\, \beta $ are real parameters
restricted to
$$ 0<\alpha\le 2\,,\q |\theta| \le \hbox{min} \{\alpha, 2-\alpha\}\,,
  \quad 0<\beta \le 1\,.\eqno(2.3)$$
In (2.2)
$\,_xD_\theta^\alpha \,$ is
the 
{\it Riesz-Feller fractional derivative}  
(in space) of order $\alpha $ and skewness $\theta\,,$
and  $\,_tD_*^\beta\,$	is
 the  {\it Caputo fractional derivative} (in time) of order $\beta \,.$
\vsp
The definitions of these fractional derivatives
are more easily understood if given
in terms of Fourier  and Laplace transforms, respectively.
\vsp
In terms of the Fourier transform 
the {\it Riesz-Feller fractional derivative} in space
is defined as 
$$ {\cal F} \l\{\, _xD_\theta^\alpha\, f(x);\kappa \r\} =
  - \psi_\theta^\alpha(\kappa ) \,
  \, \widehat f(\kappa) \,,
\q
   \psi_\theta^\alpha(\kappa ) =
|\kappa|^\alpha \, \e^{\ds  i (\sgn \kappa)\theta\pi/2}\,,
\eqno(2.4)$$
 where
$  \widehat f(\kappa)  =
{\cal F} \l\{ f(x);\kappa \r\}
  = \int_{-\infty}^{+\infty} \e^{\,\ds +i\kappa x}\,f(x)\, dx\,.$
In other words 
$\,_xD_\theta^\alpha$ is a pseudo-differential operator
\footnote{
Let us recall that a generic pseudo-differential operator $A$,
acting with
respect to the variable $x \in \RR$, is defined through its Fourier
representation, namely
$\widehat{A\,f}(\kappa) =
\widehat A(\kappa )\,\widehat f(\kappa)$
where $\widehat A(\kappa)\,$ is referred to as the symbol of $A\,.$
The $n$-th derivative operator
$\,_xD^n={d^n \over  dx^n}$
is a special case with symbol
$\widehat {\,_xD^n}(\kappa) = (-i\kappa )^n\,.$
Generally speaking, a pseudo-differential operator $A$
turns out to be defined through a
kernel of a space convolution integral;
this kernel is thus a sufficiently well-behaved function
(absolutely) integrable in $\RR$  which degenerates to a delta-type
distribution when $A =\,_xD^n\,. $
Furthermore, as a matter of fact, the symbol is given by the rule
$ \widehat A (\kappa ) = \l( A\, \e^{\, -i\kappa x}\r)\,
    \e^{\, +i\kappa x}\,. $}
with symbol 
$\,\widehat{_xD_\theta^\alpha}(\kappa) = - \psi_\theta^\alpha(\kappa)\,,$
which is   the logarithm of the
characteristic function of the generic	{\it strictly stable}
(in the L\'evy sense)
probability density, according to the Feller parameterization
\cite{Feller 52,Feller 71}.
\vsp
\begin{figure}
\begin{center}
 \includegraphics[width=.60\textwidth]{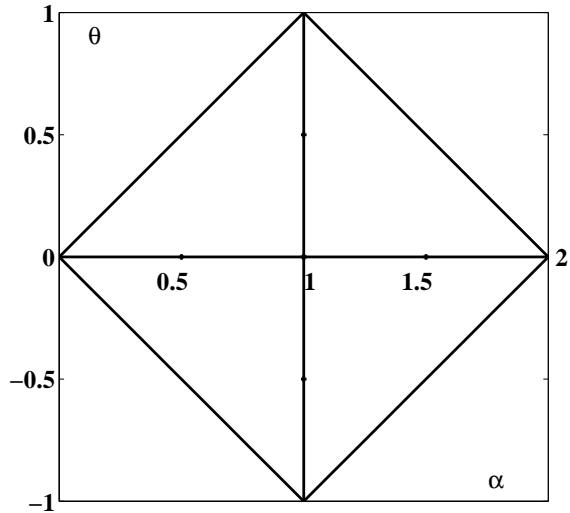}
\end{center}
\caption{The Feller-Takayasu diamond
 of the generic Riesz-Feller derivative}
\end{figure}
\vsp
 We note that the allowed region for the 
parameters $\alpha $ and $\theta$
turns out to be
 a {diamond} in the plane $\{\alpha \,,\, \theta\}$
with vertices in the points
$(0,0)\,, \,(1,1)\,,\,(2,0)\,,\, (1,-1) \,,$
that we call the {\it Feller-Takayasu diamond}
\footnote{Our notation	for the
stable distributions has been adapted
from the original one by Feller. From 1998,
see \cite{GorMai FCAA98}, we have found it as the
most convenient among the others available in the
literature, see \eg Janicki \& Weron \cite{JanickiWeron 94},
L\'evy \cite{Levy STABLE},
Montroll and associates \cite{MontrollShlesingher 84,MontrollWest 79},
Samorodnitsky \& Taqqu \cite {SamoTaqqu 94},
Sato \cite{Sato 99},
Uchaikin \& Zolotarev \cite{UchaikinZolotarev 99},
Zolotarev \cite{Zolotarev 86}.
Furthermore, this notation
has the advantage that all the class of the {\it strictly stable}
densities are represented. As far as we know,
the diamond representation in the plane
$\{\alpha ,\theta\}$ was formerly given by Takayasu
in his 1990 book on {\it Fractals}
\cite{Takayasu FRACTALS}.}, see Fig. 1.
\vsp
For $\alpha =2$ (henceforth $\theta=0$) we have
$ \widehat{\, _xD_0^2}(\kappa)  =  -\kappa ^2
= (-i\kappa )^2\,,$ so
we recover the standard
second derivative. More generally for $\theta=0$
we have
$ \widehat{\,_xD_0^\alpha} (\kappa) =
	      -|\kappa |^\alpha  = - (\kappa ^2)^{\alpha /2}$ so
$$ \,_xD_0^\alpha  = - \l(-{d^2\over dx^2}\r) ^{\alpha/2}\,.
     \eqno(2.5) $$
In this case we refer to the LHS of (2.5) as simply to the
{\it Riesz fractional derivative} of order $\alpha \,.$
We refer to the Appendix A  for the 
explicit expression of the generic Riesz-Feller derivative.
\vsp
Let us now consider the {\it Caputo fractional derivative}
in time.
Following the original idea by Caputo
 \cite{Caputo 67,Caputo 69}, see also
\cite{CaputoMaina PAGEOPH,CaputoMaina 71}, \cite{GorMai CISM97},
\cite{Podlubny 99},
a proper time fractional derivative of order $\beta \in (0,1)\,,$
useful for physical applications, may be defined in terms
of the following  rule for the Laplace transform
$$ {\cal L} \l\{ _tD_*^\beta \,f(t) ;s\r\} =
      s^\beta \,  \widetilde f(s)
   -	s^{\beta  -1}\, f(0^+) \,,
  \q 0<\beta  < 1 \,, \eqno(2.6)$$
where
$ \widetilde f(s) =
{\cal L} \l\{ f(t);s\r\}
 = \int_0^{\infty} \e^{\ds \, -st}\, f(t)\, dt\,.$
 Then
the {\it Caputo fractional derivative} of 
$f(t)$	turns out to be defined as
$$
    _tD_*^\beta \,f(t) := \cases{
       {\ds \rec{\Gamma(1-\beta )}}\,{\ds\int_0^t
 {\ds {f^{(1)}(\tau) \over (t-\tau )^\beta}\, d\tau}} \,,
 & $\; 0<\beta	<1\,,$\cr\cr
 {\ds {d f(t) \over dt}}\,, & $\; \beta =1\,.$\cr}
\eqno(2.7)$$
In other words the  operator
$\,_tD_*^\beta $ is required to generalize the well-known
rule for the Laplace transform of
the first derivative of a given (causal) function keeping the standard
initial value of the function itself.
We refer to the Appendix B for the relation
of this derivative to the more common
Riemann Liouville fractional integral and derivative.
Here we report the most relevant formula (B.6)
which provides alternative expressions for
the Caputo fractional derivative for $0<\beta <1\,:$
$$
 _tD_*^\beta \,f(t) =	 \cases{
 {\ds \rec{\Gamma(1-\beta )}\, {d \over dt}}\,
 {\ds \int_0^t {f(\tau)- f(0^+) \over (t-\tau )^\beta}\, d\tau}\,,
 \cr\cr
 {\ds \rec{\Gamma(1-\beta )}\, {d \over dt}}\,
{\ds \int_0^t  {f(\tau) \over (t-\tau )^\beta}\, d\tau}
  - f(0^+) \, {\ds {t^{-\beta} \over \Gamma(1-\beta)}} \,.\cr}
 \eqno(2.8)$$
It is worth to note that
the  time fractional derivative in the L.H.S.
of  Eq. (2.2)
can be removed by a suitable fractional integration, see (B.4),
leading
to the alternative form
$$   u(x,t) =  u(x,0^+)    +
  {\ds{1\over \Gamma(\beta )}}\,
  {\ds \int_0^t}
 {\ds \,_x D^\alpha _\theta}\,u(x,\tau )\ \,
  {d\tau \over (t-\tau)^{1-\beta}}  \,.   \eqno(2.9)$$
Differentiating  with respect to time we have another equivalent form
$$ {\d	\over \dt }  \,    u(x,t) =
  {\ds{1\over \Gamma(\beta )}}\,  {\d  \over \dt }  \,	\l\{
  {\ds \int_0^t}
 {\ds \,_x D^\alpha _\theta}\,u(x,\tau ) \,
  {d\tau \over (t-\tau)^{1-\beta}}
  \r\}\,.    \eqno(2.10)$$
It is well known that in the case of standard diffusion
Eq (2.1) can be derived
from 
the {\it continuity equation}, 
$$ {\d	\over \dt }  \,    u(x,t) +
{\ds{\d \over \dx}}\,F\l[u(x,t )\r]   =0\,, \eqno(2.11)$$
where $F$ is the {\it flux} given by
$$ F\l[u(x,t )\r]    = 
  - {\ds{\d \over \dx}}\, u(x,t)\,. \eqno(2.12)$$
Whereas Eq. (2.11) is related to a conservation  law
of a physical quantity (therefore it has a universal character), 
Eq. (2.12) is a phenomenological law which states  that
the flux is  proportional to the gradient of the field variable
(the transported quantity) 
with opposite sign. It is met in several physical contexts with
different names: \eg 
when $u$ is the temperature it is known
as Fourier's  law,  when $u$ is the pore pressure
as Darcy's law, 
when $u$ is a concentration of particles,
as Fick's law. Here we use the last  terminology  
in view of the possible applications in chemical physics.  
We recall that Fick's law is essentially an empirical law,
which represents the simplest (local in space and time) 
relationship between the flux $F$ and the gradient of 
the concentration $u$ observable in several physical phenomena.
As a matter of fact, for some experimental evidences 
in complex transport phenomena,
this law can be replaced by a more suitable phenomenological relationship
which may account for possible non-local 
and  memory effects, without
violating the conservation law 
expressed by the continuity equation (2.11).
Now it is not difficult  to derive our fractional
diffusion equation in the form (2.10) from
a generalized Fick's law in which
a suitable space-time operator depending on
$\alpha, \theta$ and $\beta$ is acting
on the gradient.   
After simple manipulations based on recent results by  
Paradisi et al \cite{Paradisi PhysA01,Paradisi PhysChem01} and
Gorenflo et al \cite {GorMai JVC02}, we can write 
$$ F(x,t) =  \, _tD^{1-\beta}\;
   _xP_\theta^\alpha  \,
\l[ - {\d\over \d x} \,u(x,t)\r] \,,
\eqno(2.13)$$
where $\,_tD^{1-\beta}$ denotes the  Riemann-Liouville
fractional derivative of order $1-\beta$ (in time), see (B.1), and 
$\,_xP_\theta^\alpha$ is the pseudo-differential operator
with symbol
$$ \widehat{\,_xP_\theta^\alpha}(\kappa)  :=
 {\widehat{\,_xD_\theta^\alpha}
(\kappa)\over\widehat{\,_xD^2}(\kappa)} =
 |\kappa|^{\ds \,\alpha-2} \, \e^{\,\ds i(\sg \kappa)\theta \pi/2} \,.
  \eqno(2.14)$$
For $\alpha =2$ and $\beta=1$ we recover  the standard Fick's law
since in this case  $\,_tD^{0}\, \,_xP^2_0 = I$ (Identity).
For $1<\alpha <2\,$  Eq. (2.13)	results to be a
a non-local connection between the flux
and the concentration gradient both from temporal
and spatial view points. For the  nature of the
operators involved,  
Eq. (2.13)   can be referred  to as the {\it fractional
Fick's law}: it turns out  to be  consistent
with the space-time fractional
diffusion equation (2.2). 
We note that for $0<\alpha \le 1\,$ Eq. (2.13)
is  meaningless 
since the symbol of the pseudo-differential
$P^\alpha_\theta$ in (2.14)
exhibits at $\kappa =0$ a singularity not Fourier integrable, which means
that the kernel is not integrable in $\RR$
\footnote{
From a purely mathematical view point one could overcome the above trouble
for $0< \alpha \le 1\,$ by stating the relationship between 
the flux $F$ 
and the concentration $u$  as
$$  F(x,t) =  \,_tD^{1-\beta} \, \,_x Q_\theta^\alpha  \,
\l [ u(x,t) \r ] \,,
$$
where 
$ \,_x Q_\theta^\alpha$ is the pseudo-differential operator with symbol
$$ \widehat{\,_xQ_\theta^\alpha}(\kappa)  :=
- \widehat{\,_xP_\theta^\alpha}(\kappa)\, \widehat{\, _xD^1}(\kappa) =
 |\kappa|^{\ds \,\alpha-1} \, \e^{\,\ds i(\sg \kappa)(\theta+1) \pi/2} \,.
  $$
 Physical reasons, however, lead us to avoid the range $0 <\alpha \le 1\,.$
In fact, for this range,  $\widehat{\,_xQ_\theta^\alpha}(\kappa)$
would be a decreasing (or constant) function of $|\kappa|\,, $ which
means that the contribution to the flux of the larger scales would
be greater than (or equal to) that of the smaller ones, which is
meaningless.}.
Generalized Fick's laws with fractional derivatives have also been 
considered by other authors, including \eg Zanette \cite{Zanette 98}
and Caputo \cite{Caputo 99,Caputo 00}. 
\section{The Green function for the space-time fractional diffusion}

The fundamental solution (or the {\it Green function})
for the space-time fractional diffusion
is intended to be the solution	of the governing equation
 (2.2), or  (2.9) or  (2.10),
 corresponding to the initial condition
$u(x,0^+) = \delta (x)\,,$
It will be denoted by $\G(x,t)\,.$
In the case of	standard diffusion, see Eq. (2.1),
the Green function is nothing but
the Gaussian probability density function with variance
$\sigma^2  =2t\,,$ namely
$$G_{2,1}^0 (x,t)
 = {1\over 2\sqrt{\pi }}\,t^{-1/2}\, \e^{-\ds x^2/(4t)}\,.
\eqno(3.1)$$
In the general case,
following the arguments by Mainardi, Luchko \& Pagnini
\cite{Mainardi LUMAPA01}
we can prove that $\G(x,t)$ is still a probability
density evolving in time with the noteworthy scaling property
$$ \G(x,t)  =
    t^{-\beta /\alpha}\,\K \l(x/t^{\beta/\alpha}\r)\,.
   \eqno(3.2)	   $$
Here $ x/t^{\beta/\alpha}\,$ acts as the similarity variable  and
$\K (\cdot)$ as  the {\it reduced Green function}.
For the analytical and computational determination of
the  reduced Green function, from now on
we  restrict our attention  to $x>0\,$
because of the
{\it symmetry relation} 
$ K_{\alpha ,\beta}^\theta(-x) = K_{\alpha ,\beta}^{-\theta}(x)
\,.$
Mainardi, Luchko \& Pagnini \cite{Mainardi LUMAPA01} have provided
(for $x>0$) the Mellin-Barnes integral representation
\footnote{
The names refer to the two authors, who in the first 1910's
developed the theory of these integrals  using them
for a complete integration of the hypergeometric differential equation.
However, as pointed out  in \cite{Erdelyi HTF}
(Vol. 1, Ch. 1, \S 1.19, p. 49), these integrals were first used
by S. Pincherle in 1888. For a revisited analysis of the pioneering work
of Pincherle (Professor of Mathematics at the
University of Bologna from 1880 to 1928) we refer
to the recent paper by
Mainardi and Pagnini \cite{MainardiPagnini OPSFA01}.
As a matter of fact this type of integral turns out to be useful
in inverting the Mellin transform, as shown hereafter.
If
$$
   {\cal M} \, \{ f(r ); s\} = f^*(s)=
   \int_0^{+\infty} f(r)\,
 r^{s-1}\,  dr,  \q  \gamma_1< \Re\, (s) <\gamma_2
$$
denotes the Mellin transform of a sufficiently well-behaved function
$f(r)\,,$ the inversion is provided by
$$
 {\cal M}^{-1}\, \{  f^*(s ); r \} =f(r)=
{1\over 2\pi i}\int_{\gamma -i
\infty}^{\gamma +i\infty} f^*(s)\, r^{-s} \,ds
$$
where $\ r>0\,,$ $\, \gamma = \Re\,(s) \,,$
$\, \gamma_1< \gamma <\gamma_2\,.$}
$$  \K(x) =
{1\over  \alpha x}
{1\over 2\pi i} \int_{\gamma-i\infty}^{\gamma+i\infty}
{\Gamma({s\over \alpha}) \, \Gamma(1-{s\over \alpha}) \,\Gamma(1-s)
 \over \Gamma(1-{\beta\over \alpha}s) \,
 \Gamma ( \rho \,s)\,
 \Gamma (1-\rho \,s)}
 \, x^{\,\ds s}\,  ds
\,,\; \rho =   { \alpha -\theta \over 2\,\alpha }\,,
\eqno(3.3)  $$
where $0< \gamma < \hbox{min} \{\alpha ,1\}\,. $

We recognize from the footnote $\null^7$
that Eq. (3.3) by changing $s$
into $-s$ can be interpreted as a Mellin transform pair
that allows us to  write the Mellin transform of
$x \, K_{\alpha ,\beta}^\theta(x)$ as $$ \q\q\q \q\q
	     \int_0^{+\infty}	   \!\!
 K_{\alpha ,\beta}^\theta(x) \,x^{\,\ds s} \,dx =
 \rec{\alpha }\,
{\Gamma(-{s/ \alpha}) \, \Gamma(1+{s/ \alpha}) \,\Gamma(1+s)
 \over \Gamma(1+{\beta s/ \alpha}) \,
 \Gamma (-  \rho \,s)\,
 \Gamma (1+ \rho \,s)} \,,\q\q\q\q \eqno(3.4)$$
$$   -\hbox{min}\{\alpha, 1\}< \Re (s) <0\,.$$
In order to include $s =0$ in the convergence strip (so, in particular, the
integral of $\K(x)$ in $\RR_0^+$ can be evaluated)
we properly use in (3.4) the functional equation
$\Gamma(1+z) = z\,\Gamma(z)$
to obtain
$$ \q\q\q \q\q \int_0^{+\infty}  \!\!
 K_{\alpha ,\beta}^\theta(x) \,x^{\, \ds s} \,dx =
 \rho \,
 { \Gamma(1-{s/\alpha})\,\Gamma(1+{s/\alpha}) \,\Gamma(1+s)
 \over
 \Gamma (1-\rho \,s)\, \Gamma (1+\rho \,s)\,\Gamma(1+{\beta\,s/ \alpha})}
\,,\q\q \q\q \q  \eqno(3.5) $$
$$ -\hbox{min}\{\alpha, 1\}< \Re (s) <\alpha \,. $$
In particular 
we find
$  \int_0^{+\infty}   K_{\alpha ,\beta}^\theta(x)  \,dx = \rho \,$
(with $ \rho  = 1/2$ if $\theta =0$).
\vsp
We note that Eq. (3.5) is strictly valid as soon as cancellations in the
"gamma fraction" at the RHS   are not possible.
Then this equation  allows us to evaluate (in $\RR_0^+$)
the (absolute) moments of order $\delta  $ for the Green function
such that   $ -\hbox{min}\{\alpha, 1\} <\delta	<\alpha\,.  $
In other words, it states that
$\,K_{\alpha ,\beta}^\theta(x)= {\cal O} \l(x^{-(\alpha +1)}\r)$
 as $x \to +\infty\,.$
When cancellations occur  in the "gamma fraction"  the range
of $\delta$ may change. 
An interesting case is 
$\,\{ \alpha =2, \, \theta =0, \, 0<\beta \le 1\}$
 ({\it time-fractional diffusion} including {\it standard diffusion}),
where Eq. (3.5)
reduces to $$ \int_0^{+\infty}	\!\!
 K_{2,\beta}^0(x) \,x^{\, \ds s} \,dx =
 \rec{2} \,
 {\Gamma(1+s)
 \over
\Gamma(1+{\beta\,s/2})}
\,,\q \Re (s) >-1\,.
 \eqno(3.6) $$
This result  proves
the existence of all moments of order $\delta  >-1$
for the corresponding Green function. In virtue of
(3.2), (3.6) 
we have
$$ \int_{-\infty}^{+\infty} \!\!\!  |x|^{\ds \,\delta }
\, G_{2,\beta}^0 (x,t)\,dx
=    {\Gamma(\delta +1)\over \Gamma(\beta \,\delta/2 +1)}\,
 t^{\ds\, \beta \,\delta /2 }\,, \q \delta >-1\,,
 \q 0<\beta \le 1 \,,  \eqno(3.7)$$
and,  for $\delta =2\,,$ the following formula
for the variance
$$ \int_{-\infty}^{+\infty} \!\!\!  x^{\ds \,2 }
\, G_{2,\beta}^0 (x,t)\,dx
=    {2\over \Gamma(\beta +1)}\,
 t^{\ds\, \beta }\,,
 \q 0<\beta \le 1 \,.  \eqno(3.8)$$
 We note that the
 Mellin-Barnes integral representation allows us to
 construct computationally   the fundamental solutions
of Eq. (2.1) for any triplet $\{\alpha ,\beta, \theta\}$
by  matching their
convergent and asymptotic expansions, see
 \cite{Mainardi LUMAPA01}, \cite{Pagnini THESIS00}.
Readers acquainted with  Fox $H$ functions can recognize
in (3.3) the representation of a certain function of this
class, see \eg
\cite{Hilfer 00a}, \cite{MathaiSaxena H},
\cite{Metzler PhysA94},
\cite{SKM 93},
\cite{Schneider LNP86}, \cite{SchneiderWyss 89}
\cite{Srivastava H}, \cite{UchaikinZolotarev 99}.
Unfortunately, as far as we know, computing routines for this general class
of special functions are not yet available.

Let us now point out the main characteristics of the peculiar cases
of {\it strictly space fractional diffusion},
{\it strictly time fractional diffusion},
and {\it neutral fractional diffusion}
based on the results stated in \cite{Mainardi LUMAPA01}.
\vsp
For  $\beta =1$ and $0<\alpha <2$
({\it strictly space fractional diffusion})
we recover
the class of the 
strictly stable (non-Gaussian)
densities exhibiting heavy  tails (with the algebraic decay
$\propto |x|^{-(\alpha +1)}$)
  and infinite variance.
For $\alpha =2$ and $0<\beta <1$
({\it strictly time fractional diffusion})
we recover
the class
of the Wright-type densities exhibiting stretched exponential
tails	 and finite variance proportional to $t^\beta \,.$
Mathematical details on these two classes	of
probability densities
can be found   
in \cite{Mainardi LUMAPA01};  for further reading 
we refer   to Schneider \cite{Schneider LNP86} for stable densities, 
and to Gorenflo, Luchko \& Mainardi \cite{GoLuMa 99,GoLuMa 00}
for the Wright-type densities.
\vsp
As for the  stochastic processes
governed by  these distributions we can expect the
following.
\vsp
For the  case of  non-Gaussian stable densities
we expect  a special class of Markovian processes, called
stable L\'evy motions, which exhibit infinite variance
associated to the possibility of arbitrarily large jumps
({\it L\'evy flights}), whereas for the case   of Wright-type densities
we expect a  class of stochastic non-Markovian processes, which
 exhibit a (finite) variance consistent with slow anomalous diffusion.
\vsp
For the special case $\alpha =\beta \le 1$ ({\it neutral diffusion})
we obtain from (3.3) an elementary  (non-negative) expression
$$ \qq \qq \qq K_{\alpha,\alpha}^\theta (x) = 
{1\over  \alpha x}
{1\over 2\pi i} \int_{\gamma-i\infty}^{\gamma+i\infty}
{\Gamma({s\over \alpha}) \, \Gamma(1-{s\over \alpha})
 \over
 \Gamma ( \rho \,s)\,
 \Gamma (1-\rho \,s)}
 \, x^{\,\ds s}\,  ds \qq  \qq \qq
\eqno(3.9)$$
$$  =
{1\over  \alpha x}
{1\over 2\pi i} \int_{\gamma-i\infty}^{\gamma+i\infty}
 {\sin (\pi\, \rho \,s)
 \over
 \sin (\pi \,s/\alpha )}
 \, x^{\,\ds s}\,  ds
=     {{1\over\pi}}\,{x^{\alpha-1} \sin[{\pi\over 2}(\alpha -\theta )] \over
1 + 2x^\alpha \cos[{\pi\over 2}(\alpha -\theta)] + x^{2\alpha}}
\,, \q x>0\,,
     $$
where $0<\gamma <\alpha \,. $
\vsp
For the generic case of {\it strictly space-time diffusion}
($0<\alpha <2, \, 0<\beta <1$), 
including neutral diffusion for $\alpha = \beta <1$,
Mainardi, Luchko \& Pagnini \cite{Mainardi LUMAPA01}
have proven the non negativity of the corresponding
reduced Green function and consequently its 
interpretation as probability density.
In this case we obtain  a class of  probability densities
(symmetric or non-symmetric according to $\theta =0$ or $\theta \ne 0$)
which exhibit  heavy tails
with an algebraic decay $\propto |x|^{-(\alpha +1)}\,.$
Thus they belong to the domain of attraction of the L\'evy stable densities
of index $\alpha $ and	can be referred to as
{\it fractional stable densities}, according to
a terminology proposed by Uchaikin \cite{Uchaikin PC00}.
The related stochastic processes
are expected to possess the  characteristics of the previous two classes.
Indeed,  they are non-Markovian (being $\beta < 1$)
and exhibit infinite variance
associated to the possibility of arbitrarily large jumps
(being $\alpha <2$).

\section{The discrete random walk models for the
 Markovian  fractional diffusion}

It is  known that a numerical approach
to the standard diffusion equation (2.1)
based on a proper finite difference
scheme	provides a discrete  Markovian
random walk model for the classical Brownian motion,
see \eg Zauderer \cite{Zauderer 89}.
\vsp
In this section we intend to generalize this approach
(that will be hereafter recalled)
in order to
provide  a discrete  Markovian
random walk model for the L\'evy  stable motion
of any order $\alpha \in (0,2)$ and skewness $\theta$ restricted
as in (2.3).  For this purpose we
present a notable finite-difference approach to
the strictly space  fractional diffusion equation
subjected to relevant restrictions, as we shall show in the following.
\vsp
The common starting point of our analysis is obviously the discretization
of the space-time  domain
 by grid points   and  time instants as follows
$$ \cases{x_j = j\, h\,,  \q h>0\,, \q j =0, \pm 1,\pm 2, \dots\,;\cr
 t_n = n \, \tau \,, \q \tau >0 \,,\q	n= 0,1,2, \dots\cr}  \eqno(4.1)
$$
where the steps $h$ and $\tau $ are assumed to be small enough.
The dependent variable $u$ is then discretized
by introducing
$y_j(t_n)$ as 
$$
  y_j(t_n) \approx
\int_{x_j-h/2}^{x_j+h/2}\!\!\! u(x,t_n) \, dx
     \approx  h\, u(x_j,t_n)\,.\eqno(4.2)$$
\subsection{The discrete random walk model for the standard diffusion}

Let us now  consider the standard diffusion equation (2.1).
With the quantities  $y_j(t_n)$ so intended,
we replace Eq.
(2.1),
after multiplication by the spatial mesh-width $h$,
 by the finite-difference equation
$$
{y_j(t_{n+1}) -y_j(t_n)\over \tau} =  {
 y_{j+1}(t_n)- 2 y_{j}(t_n) + y_{j-1}(t_n)\over h^2}\,.
   \eqno(4.3)$$
 accepting that for positive $n$ in (4.3) we have approximate
 instead of exact equality.
Since we are interested to approximate the fundamental solution (the Green
function), we must equip (4.3) with the initial condition
$ y_j(0) = \delta_ {j \, 0}\,,$
where the Kronecker symbol represents  the discrete counterpart of
the Dirac delta function.
\vsp
This approach can be interpreted as a discrete (in space and time)
{\it redistribution process} of some extensive quantity
provided it is conservative.
If  the extensive quantity is non-negative, \eg mass
or   a sojourn	probability, we have to preserve
its non-negativity.
In the first case the  $y_j(t_n)$  are imagined as
clumps of mass,   sitting at
grid points $x=x_j$ in instants $t=t_n\,, $
which collect approximatively the
total mass in the interval
$ x_j-h/2 < x \le x_j +h/2\,.$
In the second case,
the $y_j(t_n)$	may be interpreted as the probability of
sojourn in  point $x_j$  at time $t_n\, $ for a particle
making a {\it random walk} on the spatial grid in discrete instants.
 From now on, we agree to pursue the probabilistic point of view.
\vsp
In order to
have  a {\it conservative} and
{\it non-negativity preserving}  redistribution process,
the discrete variable $y_j$ is subjected to the conditions
 $$   
 \sum_{j= -\infty}^{+\infty} y_j(t_n) =
  \sum_{j= -\infty}^{+\infty} y_j(0)  \,,
 \q  y_j(t_n) \ge 0 \,, \q \hbox{for}\;\; \hbox{all} \; j \in \ZZ\,,\;
   n\in \NN_0\,.\eqno(4.4) $$
We easily recognize that our discrete redistribution process is
akin to a {\it Markov chain}:
when time proceeds from $t=t_n$
to $t=t_{n+1}\,, $  the sojourn-probabilities
are redistributed  according to the {\it transition law}
$$ y_j(t_{n+1}) = \sum _{k=-\infty }^{\infty }\,p_k\,y_{j-k}(t_n) \,,\q
j\in\ZZ\,,  \q n\in\NN_0 \,, \eqno(4.5) $$
where the $p_k$ denote suitable  {\it transfer coefficients},
which represent the probability of transition
from $x_{j-k}$ to $x_j$ (likewise from $x_j$ to $x_{j+k}$).
The transfer coefficients are to be found consistently with  the
finite-difference
equation  (4.3) equipped with the proper initial condition.
The process turns out to be both spatially homogeneous
(the probability $p_k$ of jumping from a point $x_j$ to a
point $x_{j+k}$ not  depending on $j$)
and time-stationary (the $p_k$ not depending
on $n$), as is advised when considering our Cauchy problem
and the definition of the difference operators.
Furthermore the transfer coefficients
must satisfy the conditions
$$
\sum_{k=-\infty}^\infty p_k = 1 \,,  \q p_k\ge 0 \,,
\q k = 0\,,\, \pm 1\,,\, \pm 2\, ,\, \dots
\eqno(4.6)$$
     \vsp
The transfer coefficients
in our special case are easily deduced from (4.3) and (4.5):
they turn out to be
$$
 p_0 = 1 - 2 {\tau \over h^2}\,, \q
 p_{\pm 1} =  {\tau \over h^2}\,, \q
 p_{\pm k} = 0\,, \q k = 2,3 \dots\,,. \eqno(4.7)$$
subject to the condition
$$   0< \mu := {\tau \over h^2}\le \mu _{max} = \rec{2}\,.\eqno(4.8)$$
We refer to Eqs (4.6)-(4.8) as the basic equations
for the standard random walk model for
the Gaussian process. The constant $\mu  $ will be denoted
as the {\it scaling parameter} of the standard
diffusion equation.
We recognize that only jumps of one step to the right or  to the left
or jumps of width zero occur. This corresponds to  a well-known simple
approximate realization of the {\it Brownian motion}.
\vsp
The finite difference scheme (4.3) can
 be used
 for producing {\it sample paths} of individual particles
      performing the  Brownian motion
and for producing {\it histograms}
     of the approximate realization of the Gaussian density,
 by simulating many individual paths with the same number of time steps
 and making statistics of the final positions of the particles.
\vsp
Our simulations, based on 10,000 
realizations, have been limited to the interval $|x| \le 5\,,  $ where we
have  considered the random walks to take place,
have produced an
 histogram for the Gaussian
density at $t=1\,,$
see Fig 2 (left plate).
Then, particles leaving
this space have been ignored.
In Fig 2 (right plate) we have displayed a particular sample path
or trajectory (up) and the corresponding increment series (below) obtained
with $N= 500$ time steps, by taking an intermediate value for
the scaling factor ($\mu = 0.4 < 1/2$) and reasonable  values for the
space and time steps ($h = 0.0707\,,$ $\, \tau = 0.002$)
in order to get $N \tau  = t = 1\,. $
Of course the histogram has been depicted by adopting a space
step much larger than for the sample path, namely $h= 0.25\,. $
\vsp
\begin{figure}
\begin{center}
 \includegraphics[width=.48\textwidth]{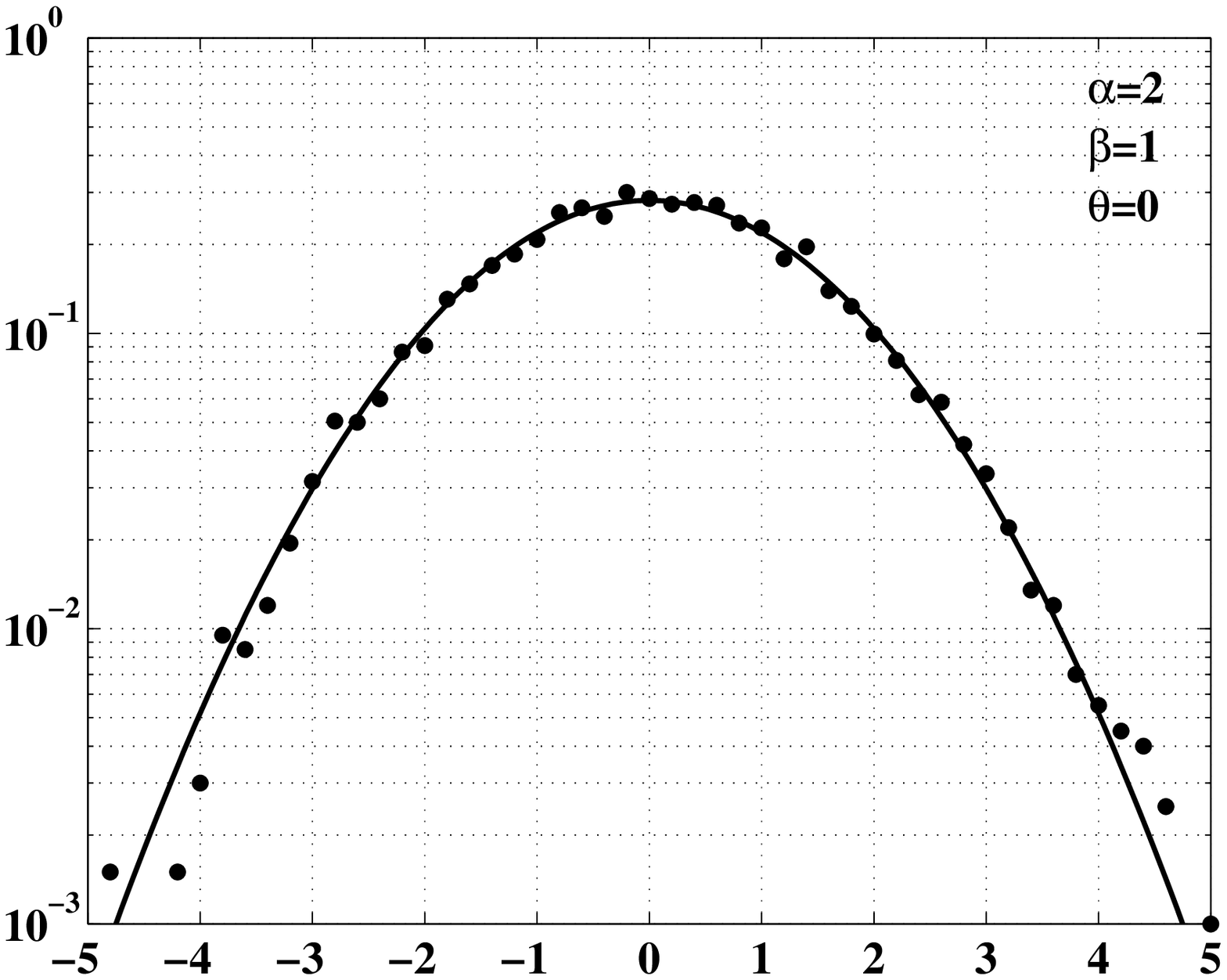}
 \includegraphics[width=.48\textwidth]{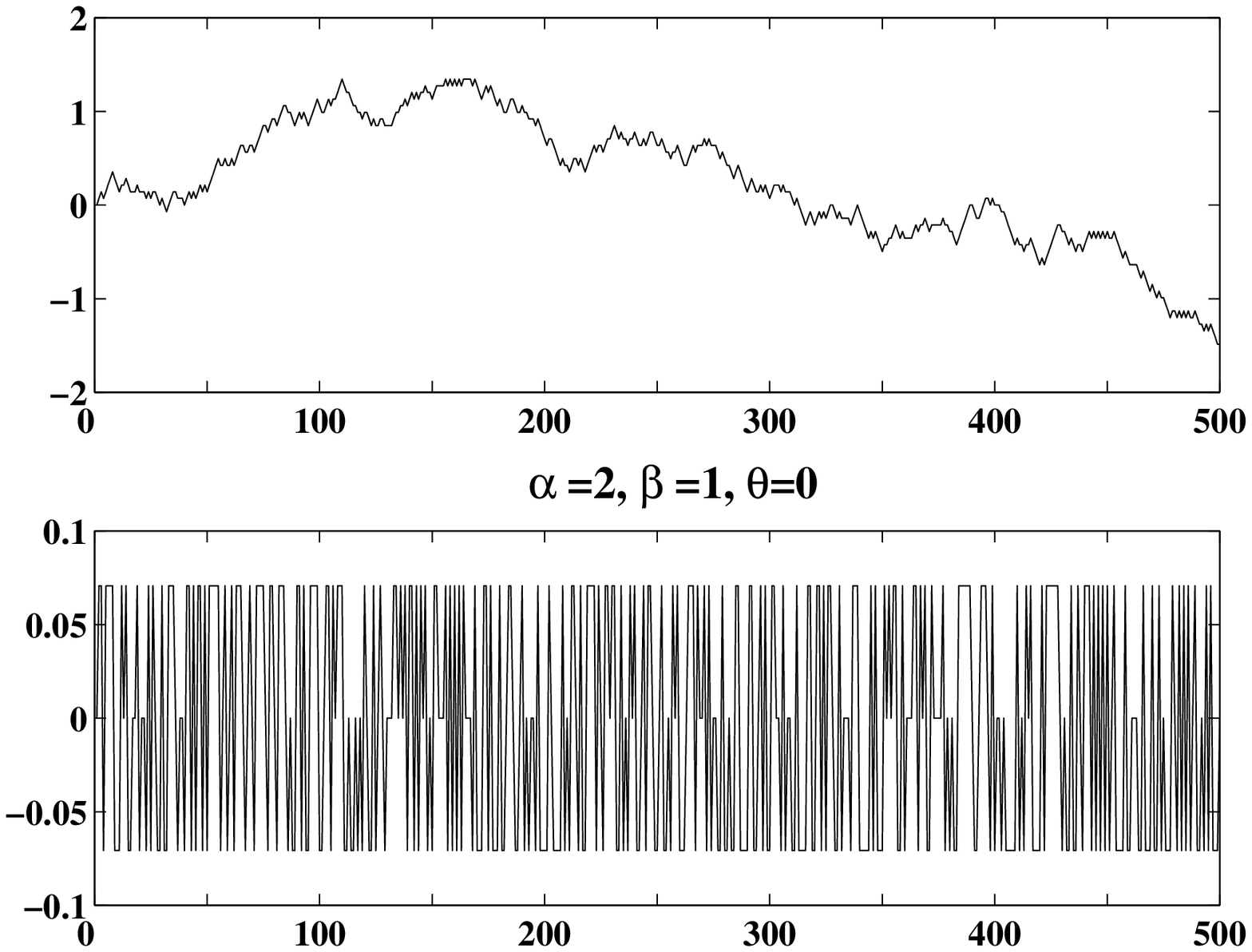}
\end{center}
\caption{Histogram (left) and a sample path with increments
(right) for standard diffusion $\{\alpha=2\,,\; \beta=1\,,\;\theta=0\}$
(Brownian motion)}
\end{figure}
\subsection{A discrete random walk model for the strictly  space fractional
diffusion}

Discretizing all the variables	as done  for the standard
diffusion equation, namely introducing a space-time mesh of widths
$h$ and $\tau$ and a discrete variable $y_j(t_n)$ interpreted as in (4.2),
the {\it essential idea}
is to replace	the {\it strictly space fractional diffusion} equation	by
the finite-difference equation
$$ {y_j(t_{n+1}) -y_j(t_n)\over \tau} \,=   \,
    _hD_\theta^{\alpha}\, y_j(t_n) \,, \q 0<\alpha < 2\,,
   |\theta| \le  \,\hbox{min}\, \{\alpha ,2-\alpha \}\,,
 \eqno(4.9)$$
where the difference operator $_hD_\theta ^{\alpha }$ is intended
to  converge to $D_\theta^{\alpha}$  (discussed
in Appendix A) as $h\to 0\,. $
As usual, we have adopted
a forward difference quotient in time at level $t=t_n$
for approximating  the first-order time derivative.
For  $_hD_\theta ^{\alpha }$
we require a scheme which must reduce as $\alpha  =2$ to
a symmetric second-order difference quotient
in space at level $t=t_n\,, $ which has been adopted for
approximating the second-order
space derivative.
 Furthermore, the
the finite-difference equation (4.9)
  is required
to be consistent with a {\it conservative}, {\it non-negativity preserving
redistribution process},
subject to the initial condition
 $y_j(0)= \delta_{j\,0}\,. $
Referring  to Eqs (A.4), (A.8) and (A.10)-(A.11),
we write
$$
_hD_\theta^{\alpha}\, y_j(t_n)	 =   
      \cases{
  -\, \left[  c_+(\alpha,\theta)\,_h D^{\alpha}_+
+ c_-(\alpha,\theta)\,_hD^{\alpha}_-  \right]\, y_j(t_n)  \,,
   &  $ \alpha \ne 1\,, $ \cr\cr
  \left[ \cos (\theta \pi/2) \,_h D_0^1 + \,\sin (\theta \pi/2) \,_h D_\pm
\right]\,	       y_j(t_n) \,,
   &  $ \alpha =1 \,.$\cr}
  \eqno(4.10)$$
With the notation $\,_h D_\pm$ we intend to adopt
$\,_h D_+$ when     $0<\theta \le 1$ and $\,_h D_-$
when	 $-1 \le \theta <0\,. $
\vsp
We thus must find
suitable finite-difference schemes
for the pseudo-differential operators entering the
Feller derivative, namely
 $_hD_\pm^{\alpha}$ (Weyl fractional derivatives of order $\alpha $) if
$\alpha \ne 1$
and $ _hD_0^1$ (Riesz derivative of first-order)
if $\alpha =1\,. $
For this purpose  
we must keep distinct the three cases
$$  \cases{\hbox{(a)} \q 0 <\alpha <1\,, & $ |\theta | \le \alpha \,;$
   \cr
    \hbox{(b)} \q 1 <\alpha < 2\,, & $ |\theta | \le 2-\alpha \,;$
 \cr
    \hbox{(c)} \q \alpha =1\,, & $ |\theta| \le 1 \,.$
\cr} \eqno(4.11)
 $$
For  $\alpha \ne 1$ the starting point
is the	Gr\"unwald-Letnikov discretization of fractional derivatives,
on which the
reader can inform himself from the treatises
on fractional calculus, see \eg
\cite{MillerRoss 93},
\cite{OldhamSpanier 74},
\cite{Podlubny 99},
\cite{SKM 93},
or in the  review article by Gorenflo \cite{Gorenflo CISM97}.
However, for our purposes, we  must make a clever use of
the  Gr\"unwald-Letnikov scheme treating separately  the two cases
(a) and (b) in order to be consistent with a
conservative, non-negativity preserving  redistribution process.
We have
$$
_h D_\pm^{\alpha} \,y_j =
\cases{
 {\ds {\!1\over h^{\alpha}}}\,\sum\limits_{k=0}^\infty (-1)^k
{\ds\l({\alpha \atop k}\r)}\, y_{j\mp k}\,,  
& in the case (a)  $0<\alpha <1\,,$
\cr
 {\ds {\!1\over h^{\alpha}}}\, \sum\limits_{k=0}^\infty (-1)^k
{\ds \l({\alpha \atop k}\r)}\, y_{j\pm 1\mp k}\,,
& in the case (b)    $1<\alpha < 2\,.$
 \cr}
 \eqno(4.12)
$$
Notice the shift of index in the case (b) which among other things has the
effect that in the limiting
case $\alpha=2$ (the classical diffusion equation) we
obtain the standard symmetric three-point difference scheme.
For more details and discussions see Gorenflo \& Mainardi
\cite{GorMai FCAA98}.
\vsp
For $\alpha =1$
we limit ourselves to consider the case $\theta =0$,
that requires
the discretization of the Hilbert transform in (A.11).
The hyper-singular   integral representation of the symmetric
space-fractional  derivative  (the "Riesz derivative")
given by   Gorenflo \& Mainardi in \cite{GorMai CHEMNITZ01}
[as formula (2.20)] readily offers us the discretization
$$ _h D_0^1 y_j =  {1 \over \pi\, h}\,
\sum_{k=1}^{\infty}{ y_{j+k} - 2 y_j + y_{j-k} \over k(k+1)}\,.
\eqno(4.13)$$
For $0 <|\theta| \le 1$ we have analogously to (A.10) to take
into account	 the forward/backward difference quotients
for $\,_h D_\pm$   
$$ _hD_+ =  {y_{j+1} - y_{j}\over h}\,, \q
   _hD_- =  {y_{j} - y_{j-1}\over h}\,,\eqno(4.14) $$
to obtain a non-negative scheme in a proper way.
However, we here leave out these "drifts" (in negative or  positive
directions)  from the computation
of the transfer coefficients $p_k$.
\vsp
Then, inserting the expressions (4.12)-(4.13) in the
finite-difference equation (4.9)-(4.10) yields the {\it transition law}
$$ y_j(t_{n+1}) = \sum _{k=-\infty }^{\infty }\,p_k\,y_{j-k}(t_n) \,,\q
j\in\ZZ\,,  \q n\in\NN_0 \,, \eqno(4.15) $$
where  the   {\it transfer coefficients} $p_k$ turn out to be,
respectively,
 $$ \cases{
p_0 = 1- \mu \, (c_+ + c_-)   = 1- \mu \,
      {\ds {\cos\,(\theta \pi/2) \over \cos\,(\alpha\pi/2)}} \,,  \cr
p_{\pm k} =
    (-1)^{k+1} \mu \, {\ds {\alpha \choose k}} \, c_\pm \,,
\q  k =1,2,\dots  \cr}	 \q 0<\alpha <1\,,\;  |\theta| \le \alpha
\,;\eqno(4.16a)$$
$$\cases{
 p_0 = 1+ \mu \, {\ds {\alpha\choose 1}}\, (c_+ + c_-)
   = 1-\mu \, \alpha \,
 {\ds {\cos\,(\theta \pi/2) \over |\cos\,(\alpha\pi/2)|}} \,,  \cr
 p_{\pm 1}=
 - \mu \,\l[{\ds{\alpha\choose 2}}\, c_\pm+c_\mp\r]
   \,,\qq 1<\alpha < 2\,,\;
      |\theta|\le 2-\alpha \,;
\cr
 p_{\pm k} =
(-1)^{k} \,\mu \, {\ds {\alpha \choose k+1}} \, c_\pm\,,
   \q  k = 2,3, \dots  \cr}   \eqno(4.16b)$$
$$ \cases{
 p_0 = 1 - {\ds {2\mu \over \pi}} \,,\cr
  p_{\pm k} =  {\ds {\mu \over \pi}} \, {\ds {1\over k(k+1)}} \,,
  \q  k=1, 2,\dots\,, \cr}
 \q \alpha =1  \,,\; \theta =0\,; \eqno(4.16c) $$
with   the scale parameter
$$  \mu:= {\tau \over h^\alpha }\,, \q 0<\alpha < 2\,. \eqno(4.17)$$
It is straightforward to check in all cases the summation condition
${\ds \sum_{k=-\infty}^\infty p_k =1\,.}$
Since all $p_{\pm k} \ge 0\,,\; k \ne 0\,$
the non-negativity condition is met if we  require $p_0 \ge 0$,
 \ie if the scale parameter $\mu $ is restricted as follows
$$
0 < \mu := {\tau \over h^\alpha } \le \mu_{max} =  \cases{
     {\ds {\cos{\alpha\pi/ 2} \over \cos{\theta\pi/2}}},
& 
 $\;0<\alpha <1\,,\; |\theta|\le \alpha\,,  $ \cr
     {\ds {1\over\alpha}{|\cos{\alpha\pi/2}|\over \cos{\theta\pi/2}}},
& 
$\;1<\alpha < 2\,,\; |\theta|\le 2- \alpha\,,$ \cr
     {\ds {\pi \over 2}},
& 
 $\;\alpha =1\,,\; \theta= 0\,.$	\cr
     } \eqno(4.18)
$$
We note that in both the limits $\alpha \to 1^-$
and   $\alpha \to 1^+$ the permissible range of the scaling factor $\mu $ is
vanishing.
In numerical practice the consequence will be that if $\alpha $ is
near 1 the convergence is slow: for good approximation
we will need a very small step-time $\tau $ with respect
to the step-length $h$
\footnote{In order to get a continuous transition to the case
 $\alpha =1$ we need to consider a different discretization scheme,
 presented  in \cite{GorDFMai PhysA99}
for the first time and rigorously analyzed in \cite{GorMai CHEMNITZ01}.
For this scheme, however, we loose the continuity as $\alpha  \to 2^-\,.$
This means that "there  is no free lunch"; we have to pay for the
good behaviour at $\alpha =1$ with bad behaviour at $\alpha =2$
in a sense described in \cite{GorDFMai PhysA99}.}.
\vsp
\begin{figure} [!h] 
\begin{center}
\includegraphics[height=7.0truecm,width=12.0truecm]{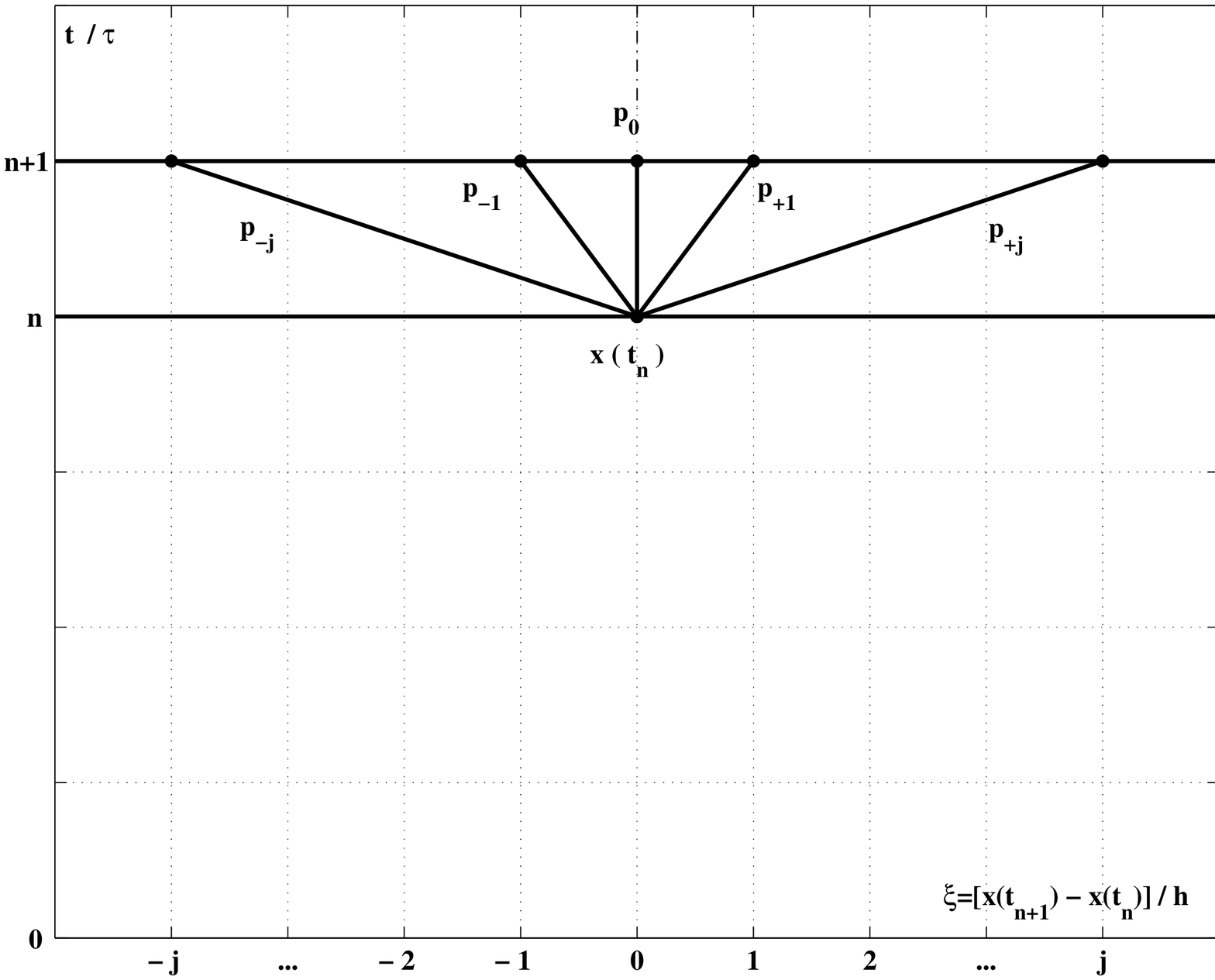}
\end{center}
\vskip -0.5truecm
 \caption{Sketch of the transition scheme for the
space-fractional random walker}
\end{figure}
\vsp
The  striking  difference    to our discretized  Brownian 
is the appearance of arbitrarily large jumps
with power-like decay probability for which  these
discrete models are referred to as {\it L\'evy flights}.
See Fig. 3 for a sketch of the transition scheme
and Section 6 for the numerical results.


\section{The discrete random walk models for the
 non-Markovian	fractional diffusion}

Let us now try to generalize the above arguments by adapting them to
the non-Markovian
 cases of our space-time fractional diffusion equation (2.2).
We find it convenient to proceed by steps: we first consider
the case  of the {\it strictly time fractional diffusion},
see also \cite{GorMai JVC02},
and then, by combining these  arguments  with those
for {\it strictly space fractional diffusion},
we treat the {\it strictly space-time fractional diffusion}.
\subsection{A discrete random walk model for the strictly time fractional
diffusion}

Discretizing all the variables as  for the standard diffusion
equation, namely introducing a space-time mesh of widths
$h$ and $\tau $ and a discrete variable $y_j(t_n)$ interpreted as in (4.2),
the {\it essential idea}
is to replace	the {\it strictly time fractional diffusion} equation
 by the
finite-difference equation
$$  _\tau D_*^{\beta} \,  y_j(t_{n+1})	=
 {y_{j+1}(t_n) - 2y_j(t_n)  + y_{j-1}(t_n)  \over h^2}
 \,,  \q 0< \beta  < 1\,,
\eqno(5.1)$$
where the difference operator $_\tau D_*^{\beta}$ is intended
to  converge to $D_*^{\beta}$  as $\tau \to 0\,. $
\newpage
\vsp
As usual, we have adopted
a symmetric second-order difference quotient
in space at level $t=t_n$ for approximating the second-order
space derivative. For  $_\tau D_*^{\beta}$
we require a scheme which must reduce as $\beta =1$ to
a forward difference quotient in time at level $t=t_n\,, $
which is usually adopted for approximating  the first-order time
derivative. Then, for approximating the time fractional derivative
(in  Caputo's sense),
we adopt a backward Gr\"unwald-Letnikov scheme in time
(starting at level $t=t_{n+1}$) which reads
\def\betak{{\beta  \choose k}}
$$  _\tau D_*^{\beta} \,  y_j(t_{n+1})	=
    \sum_{k=0}^{n+1} (-1)^k \, \betak\,
  {y_j(t_{n+1-k}) - y_j(0)\over \tau^\beta }\,,\q 0<\beta < 1
\,. \eqno(5.2)$$
Here  the subtraction of $y_j(0)$ in each term of the sum reflects the
subtraction of $f(0^+)$ in formula (B.6) for the Caputo fractional
derivative.
Combining (5.1) and (5.2), introducing the scaling
parameter
$$  \mu   := {\tau^\beta  \over  h^2}\,,\q 0<\beta \le 1\,,
 \eqno(5.3)  $$
 and using the	"empty sum" convention
$\ds {\sum_{k=p}^q =0}$ if $  q<p\,$ 
(here $p=1$ when  $q=n=0\,$),
we obtain for $n \ge 0$ ($t_0=0$):
$$ \qq
y_{j}(t_{n+1}) = y_{j}(t_0)  \,
   \sum_{k=0}^{n} (-1)^k \, \betak +
  \sum_{k=1}^{n} (-1)^{k+1} \, \betak\,   y_j(t_{n+1-k})   \qq\qq\qq
\eqno(5.4)$$
$$+ \mu \, \l[ y_{j+1}(t_n) - 2y_j(t_n)  + y_{j-1}(t_n) \r] \,.
 $$
Thus,  (5.4)  provides the  universal {\it transition law}
from $t_n$ to $t_{n+1}$  valid for all	$n \ge 0\,. $
For convenience let us introduce
the coefficients $c_k$, $b_m$
 $$ \cases{
c_k = (-1)^{k+1} \, {\ds\betak} =
 \l| {\ds \betak }\r| \,, \q  k\ge 1\,, \cr
b_m = {\ds \sum_{k=0}^m (-1)^{k} \, \betak }\,,\q m \ge 0\,. \cr}
\eqno(5.5)$$
For $\beta =1$ (standard diffusion) we note that
all these coefficients are vanishing except $b_0 = c_1 =1\,.$
For $0 < \beta < 1$ 
they possess the  properties
$$   \sum_{k=1}^\infty c_k = 1\,, \q
      1 > \beta = c_1 > c_2 > c_3 >  \,  \dots \,\to 0\,,
  \eqno(5.6)$$
$$ \cases{
   b_0=1 = {\ds \sum_{k=1}^\infty \, c_k}\,, \q b_m =1 -
       {\ds \sum_{k=1}^m \, c_k}  =
{\ds  \sum_{k=m+1}^\infty \, c_k \,,}\cr
  1=b_0 > b_1 > b_2 > b_3 > \, \dots \, \to  0\,.\cr}
 \eqno(5.7)$$
We thus observe that the $c_k$ and the $b_m$  form sequences of
positive numbers, not greater than 1,
decreasing strictly monotonically to zero.
Thanks to the introduction of the above coefficients
the universal transition law (5.4) can be written in the
following noteworthy form
$$ y_j(t_{n+1}) = b_n\, y_j(t_0) +
     \sum_{k=1}^n c_k\, y_j(t_{n+1-k})
 + \mu	\l[ y_{j+1} (t_n) -2y_j(t_n) + y_{j-1} (t_n) \r],
 \eqno(5.8)$$
with the  empty sum convention convention if $n=0\,. $
In particular we get,
for $n=0\,:$
$$ y_j(t_1) = (1-2\mu )\, y_j(t_0) + \mu \,\l[
      y_{j+1} (t_0) + y_{j-1} (t_0) \r]\,; $$
for $n =1 \,:$
$$ y_j(t_2) = b_1 y_j(t_0) + (c_1-2\mu )\, y_j(t_1) + \mu \,\l[
      y_{j+1} (t_1) + y_{j-1} (t_1) \r]\,; $$
for $n\ge 2\,:$
$$
\begin{array}{ll} 
{\ds y_j(t_{n+1})} = & {\ds b_n\, y_j(t_0) +
     \sum_{k=2}^n c_k\, y_j(t_{n+1-k})}	\\ 
  & {\ds + (c_1 -2\mu) \,y_j(t_n) + \mu
  \,\left[ y_{j+1} (t_n) + y_{j-1} (t_n) \right]}\,.
 \end{array} 
   $$
Observe that $c_1 = \beta \,. $
The scheme (5.8) {\it preserves non-negativity}, if all coefficients
are non-negative, hence if
$$ 0< \mu =
{\tau ^\beta \over h^2} \le {\beta \over 2}\,. \eqno(5.9)$$
Furthermore it is {\it conservative}, as we shall prove by
induction: \ie
$${\ds \sum_{j=-\infty}^{+\infty}}  | y_j(t_0)| <\infty\,
\Longrightarrow \,
  {\ds\sum_{j = -\infty}^{+\infty}}	 y_j(t_n)
   = {\ds\sum_{j = -\infty}^{+\infty}}	 y_j(t_0)\,, \q n \in \NN\,.
\eqno(5.10)$$
In fact,  putting
$S_n =	 {\ds\sum_{j = -\infty}^{+\infty}}  y_j(t_n) $
for $n\ge   0\,, $ then from (5.8) we get
$$
 S_1 =
  (1-2\mu )\, \sum y_j(t_0) +\mu \sum y_{j-1}(t_0)+ \mu \sum y_{j+1}(t_0)
 = S_0,,$$
and for $n\ge 1$ we find always from (5.8),
assuming $S_0 = S_1 = \dots S_n$ already proved,
$$
\begin{array}{ll}
{\ds  S_{n+1}} & 
= {\ds b_n\, S_0 + \sum_{k=2}^n c_k \, S_{n+1-k} +
 (\beta -2\mu +\mu +\mu )\, S_n} \\  
 &
 = {\ds \left( b_n +    \sum_{k=1}^n c_k \r)\, S_0 = S_0}\,, 
 \end{array}
 $$
using $\beta =c_1\,. $ We have thus proved conservativity.
Non-negativity preservation and conservativity mean that our scheme
can be interpreted as a {\it redistribution scheme} of clumps
$y_j(t_n)\,.$
For orientation on such aspects and for examples let us quote
the works by Gorenflo on  conservative difference schemes for diffusion
problems, see \eg
\cite{Gorenflo 70,Gorenflo 78}.
\vsp
The interpretation of our {\it redistribution} scheme is as follows:
the clump $y_j(t_{n+1})$ arises as a weighted-memory  average of
the (previous) $n+1$ values $y_j(t_m)\,, $ with $m = n\,,\, n-1\, ,
 \dots \,,1\,,0\,, $ with  positive  weights
$$\beta =c_1, \,c_2\,,\dots\,,	c_n\,,\q
b_n =  1- {\ds \sum_{k=1}^n c_k}\,,
\eqno(5.11)$$
followed by subtraction of $2\mu \, y_j(t_n)$, which is given in
equal parts to the    neighbouring points $x_{j-1}$
and $x_{j+1}\, $ but replaced by the contribution
$\, \mu  \,[ y_{j+1} (t_n) + y_{j-1} (t_n) ]\,$
from these neighbouring points.
For  {\it random walk} interpretation we consider the
$y_j(t_n)$ as probabilities of sojourn at point $x_j$ in instant
$t_n\, $  requiring the normalization condition
${\ds \sum_{j=-\infty}^{+\infty} y_j(t_0)}=1\,.$
\vsp
For $n=0$ Equation (5.8) means (by appropriate re-interpretation
of the spatial index $j$):
A particle sitting at $x_j$ in instant $t_0$ jumps,
when $t$ proceeds from $t_0$ to $t_1\,, $
with probability $\mu $ to the neighbour point $x_{j+1}\,, $
with probability $\mu $ to the neighbour point $x_{j-1}\,, $
and with probability $1-2\mu $ it remains at $x_{j}\,. $
For $n\ge 1$ we write (5.8), using $\beta =c_1\,,$
as follows:
$$ 
\begin{array}{ll}
{\ds y_j(t_{n+1})} = & 
{\ds \left(1- \sum_{k=1}^n c_k \right)\, y_j(t_0) +
      c_n\, y_j(t_1)  + c_{n-1}\, y_j(t_2)
      + \dots + c_2\, y_j(t_{n-1}) } \\
& {\ds + (c_1 -2\mu) \,y_j(t_n) + \mu
  \,\left[ y_{j+1} (t_n) + y_{j-1} (t_n) \right]}\,.
\end{array}
\eqno(5.12)
$$
\noindent
Obviously, all coefficients (probabilities) are non negative, and their
sum is 1.
But what does it mean? Having a particle,
sitting  in $x_j$ at instant $t_n\,, $ where will we find it
with which probability at  instant $t_{n+1}?$
From (5.12) we conclude, by re-interpretation of the spatial index $j\,, $
{\it considering the whole history} of the particle, \ie
the particle path
$ \l\{ x(t_0)\,,\,   x(t_1)\,,\, x(t_2)\,,\, \dots \,,\, x(t_n)\r\},$
 that if at instant $t_n$ it is in point $x_j\,, $
there is the contribution $c_1-2\mu $ to be again at $x_j$
at instant $t_{n+1}\,, $
the contribution $\mu $ to go to $x_{j-1}\,, $
the contribution $\mu $ to go to  $x_{j+1}\,. $
But the sum of these contributions is $c_1 =\beta \le 1\,. $
So, excluding the case $\beta =1$ in which we recover the
standard  diffusion ({\it Markovian  process}),
for $\beta <1$ we have to consider the previous
time levels ({\it non-Markovian process}).
Then, from level $t_{n-1}$
we get the contribution $c_2$ for the probability of staying in $x_j$ also
at time $t_{n+1}\,, $
from level $t_{n-2}$
we get the contribution $c_3$ for the probability of staying in $x_j$
at time $t_{n+1}\,, \dots \,,$
from level $t_{1}$
we get the contribution $c_n$ for the probability of staying in $x_j$
at time $t_{n+1}\,, $
and finally,
from level $t_{0}=0$ we get the contribution  $b_n$
for the probability of staying in $x_j$ at time $t_{n+1}\,. $
Thus, the whole history up to $t_n$	
decides probabilistically
where the particle will be at instant $t_{n+1}\,. $
\vsp
Let us consider the problem of {\it simulation} of transition
from time level $t_n$  to $t_{n+1}$:
Assume the particle sitting in $x_j$ at instant $t_n\,. $
Generate a random number equidistributed in $0\le \rho <1\,, $
and subdivide the interval $[0,1)$ as follows.
From left to right beginning at zero we put adjacent intervals
of length $c_1, c_2, \dots ,c_n, b_n$, for consistency
left-closed, right-open.  The sum of these is 1.
We divide further the first interval (of length $c_1$)
into sub-intervals of length $\mu, c_1-2\mu, \mu\,.$
Then we look into which of the above intervals
the random number falls.
If in first interval with length $c_1 = \mu + (c_1 -2\mu ) +\mu $, then
look in which subinterval, and correspondingly move the
particle to $x_{j-1}\,, $ or leave it at $x_j$ or move	 to $x_{j+1}\,.$
If the random number falls into one of the intervals
with length $c_2\,,\, c_3\,,\, \dots \,,c_n$ (\ie $c_k$ with
$2\le k \le n$), then move the particle
back to its previous position $x(t_{n+1-k})$, which by chance could
be identical with $x_j = x(t_n)\,. $
If the random number falls into the rightmost interval
with length $b_n$
then move the particle back to its initial position
$x(t_0)\,,$ for which we recommend $x(t_0) = 0\,,$
meaning $y_j(t_0) = \delta_{j 0}\,,$
 in accordance with the initial condition
      $u(x,0) = \delta(x)$  for (2.2) with $\alpha=2\,,\, \theta=0\,.$
\vsp
A sketch of the transition scheme for the random walker is
reported in Figure 4.
Besides the diffusive part $(\mu \,,\, c_1 -2\mu\,,\, \mu )$ which lets the
particle jump at most to neighbouring points, we have for $0<\beta <1$
the memory part which gives a tendency to return to former positions
even if they are far away.
Due to Equations (5.6)-(5.7), of course, the probability to return to a far
away point gets smaller and smaller the larger the time lapse is from
the instant when the particle was there.
\vsp
\begin{figure} [!h] 
\begin{center}
\includegraphics[height=7.0truecm,width=12.0truecm]{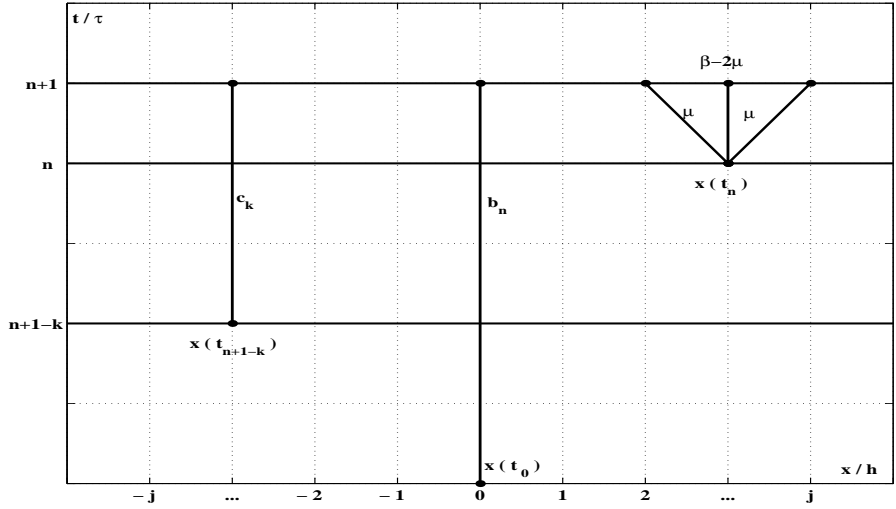}
\end{center}
\vskip -0.5truecm
 \caption{Sketch of the transition scheme for the
time-fractional random walker}
\end{figure}
\subsection{A discrete random walk model for the strictly
space-time fractional diffusion}

By  combining the approach of the preceding
Subsections 3.2 and 4.1 we can construct a discrete random walk
model for the strictly space-time fractional diffusion equation,
namely for the case$\{0<\alpha <2\,,\, 0<\beta <1\}\,.$
We replace in (2.2) the (Riesz-Feller) space-fractional
derivative by (4.10) and
the (Caputo) time-fractional  derivative by (5.2).
Solving then for the "new" value
$y_j(t_{n+1})$ we see that the scaling parameter
$\mu =\tau ^\beta /h^\alpha $ plays the essential role
in obtaining a scheme with all transition coefficients
non-negative. In fact, also here $\mu $ must be restricted to
a suitable interval $(0, \mu _{max}]\,. $
In analogy to the case of the strictly time fractional diffusion
the "discrete" diffusion in space occurs only
between the time-level $t_n$ and $t_{n+1}$
and the memory part of the process only
straight-backwards in time.
However, in contrast to the strictly time fractional
diffusion, the discrete diffusion (or the random walker)
can now go to any grid point in space,
not only to immediate neighbouring grid points.
We abstain from presenting here all the
lengthy formulas that again need distinction between
the three cases: (a) $0<\alpha <1\,,$ (b) $1<\alpha <2\,,$
and (c) $\alpha =1\,. $


\section{Numerical results and concluding discussions}

Our simulations are all based
on 10,000 realizations. In addition to the Brownian motion,
see Fig. 2, we have
considered    a set of 8 case-studies, see Figs 5-12, for our
space-time fractional diffusion that,  in our opinion,
can better illustrate  the state-of art  of this analysis.
The sample paths and the corresponding increments  are plotted against the
time steps up to 500, see the right  figure-plates, while
the histograms refer to  densities at $t=1\,$
for $|x| \le 5\,,$  see the left  figure-plates.
All the
plots were drown by using the MATLAB system.
The relevant parameters $\alpha $, $\theta$, $\beta$,
$\mu $, $h$ and $\tau $ used in the Figures
are reported in Table I.
For convenience we have also reported the maximum value
of the scaling parameter $\mu $ as can be deduced
from our theoretical analysis.
More details can be found in \cite{Moretti THESIS00}.
\begin{center}
\vskip 0.25truecm
\begin{tabular}{|c|c|c||c||c|c||c|c|}
\hline
$\alpha $ & $\theta$ & $\beta $  & $\mu_{max} $
          & $h_{S}$ & $\tau_{S}$ & $h_{H}$ & $\tau_{H}$ \\
 \hline
$2$   & $0$    & $1$   &  $0.50$
          & $7.0\,10^{-2}$ & $2.0\,10^{-3}$ & $0.20$ & $2.5\,10^{-2}$ \\
$1.75$ & $0$    & $1$   &  $0.53$
          & $4.7\,10^{-2}$ & $2.0\,10^{-3}$ & $0.33$ & $2.0\,10^{-2}$ \\
$1.75$ & $-.25$ & $1$   &  $0.57$
          & $4.5\,10^{-2}$ & $2.0\,10^{-3}$ & $0.33$ & $2.0\,10^{-2}$ \\
$1.50$ & $0$    & $1$   &  $0.47$
          & $3.0\,10^{-2}$ & $2.0\,10^{-3}$ & $0.20$ & $1.0\,10^{-2}$ \\
$1.50$ & $-.50$ & $1$   &  $0.67$
          & $2.4\,10^{-2}$ & $2.0\,10^{-3}$ & $0.20$ & $1.0\,10^{-2}$ \\
$2$    & $0$    &$0.75$ &  $0.37$
          & $0.17$ & $2.0\,10^{-3}$ & $0.25$ & $5.0\,10^{-3}$ \\
$2$    & $0$    &$0.50$ &  $0.25$
          & $0.47$ & $2.0\,10^{-3}$ & $0.50$ & $2.5\,10^{-2}$ \\
$1.50$ & $0$    &$0.50$ &  $0.24$
          & $0.38$ & $2.0\,10^{-3}$ & $0.50$ & $5.0\,10^{-3}$ \\
$1.50$ & $-.50$ &$0.50$ &  $0.33$
          & $0.30$ & $2.0\,10^{-3}$ & $0.50$ & $5.0\,10^{-3}$ \\
\hline
\end{tabular}
\\
\vskip 0.5truecm {\bf Table I}: The relevant parameters for the simulations
\\
$\alpha $ = space-fractional order,
$\, \theta$ = skewness,
$\,\beta$  = time-fractional order,
 \\
$\mu = \tau ^\beta /h^\alpha $ = scaling parameter,
\\
$ h_S $ = space-step, $\,\tau_S$ = time-step for sample paths,
\\
$ h_H $ = space-step, $\,\tau_H$ = time-step for histograms.
\end{center}
\vsp
In practice, in our numerical studies
there is required truncation for
two different causes.
As in the classical Brownian motion,
a trivial truncation is required if
a priori one wants a definite region of space to be considered
in which the walk takes place. Then, particles leaving
this space have been ignored.
However, if $0< \alpha<  2\,,$ at variance with our
discretized Brownian motion, we now
have an infinite number of transition probabilities.
Since it is impossible to simulate  all infinitely many discrete
probabilities, so the size of possible jumps must be limited to a maximal
possible jump length.
\vsp
The plates in Fig. 5  are concerning two cases 
 of {\it symmetrical, strictly space fractional diffusion}:
$\{\alpha=1.75\,,\, \beta=1,\, \theta=0\}$, 
 $\{\alpha=1.5\,,\, \beta=1,\, \theta =0\}$
whreas the plates in Fig. 6 are concerning two cases
of {\it extremal, strictly space fractional diffusion}:
$\{\alpha=1.75\,,\, \beta=1,\, \theta=-0.25\}$, 
$\{\alpha=1.5\,,\, \beta=1,\, \theta =-0.50\}$
From the sample paths in Figs. 5, 6 one can recognize the "wild" character
(with large jumps) of the {\it L\'evy flights}
with respect to the "tame" character
of the	{\it Brownian motion} 
outlined in Fig. 2.
\begin{figure}[!ht]  
\begin{center}
 \includegraphics[width=.48\textwidth]{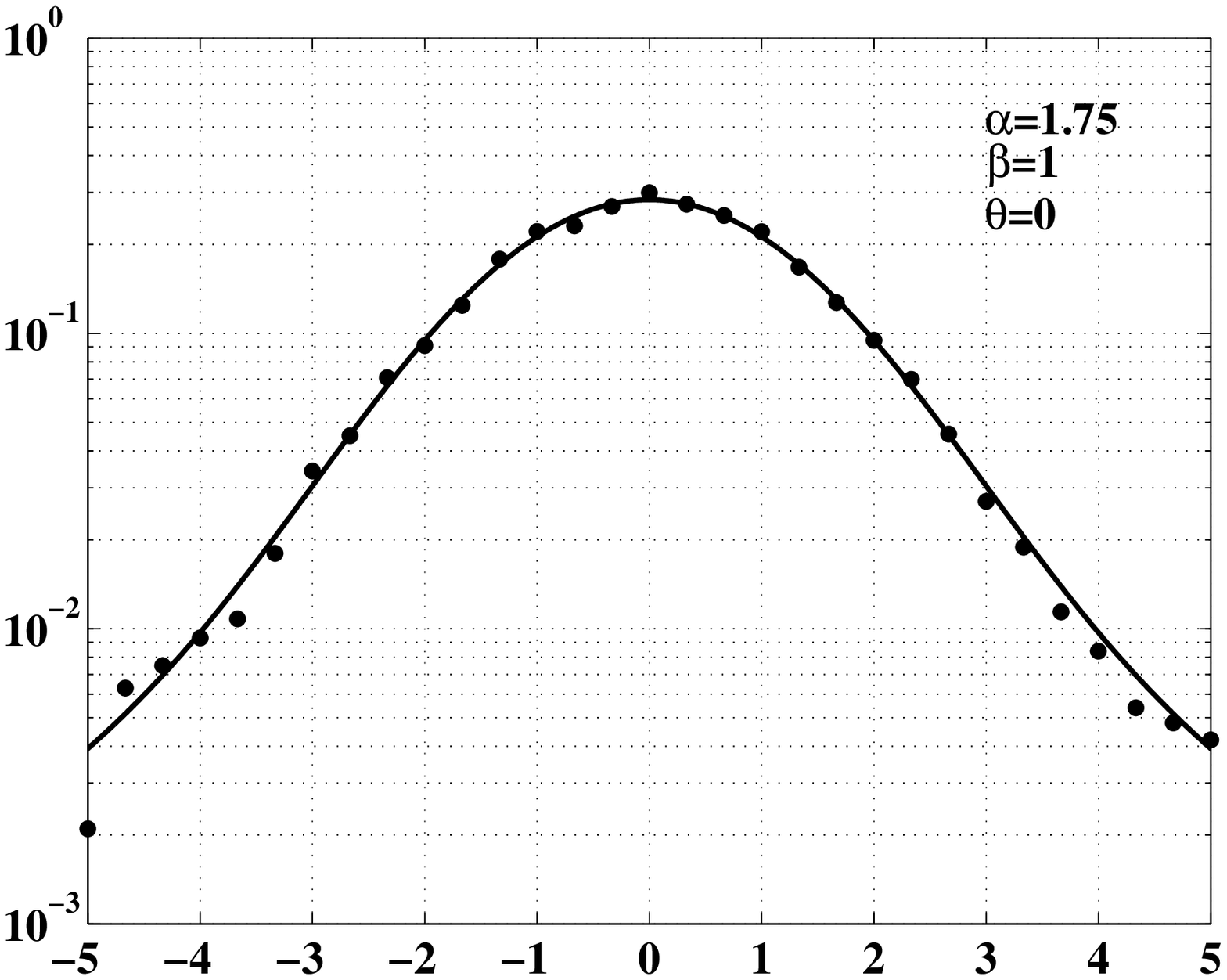}
 \includegraphics[width=.48\textwidth]{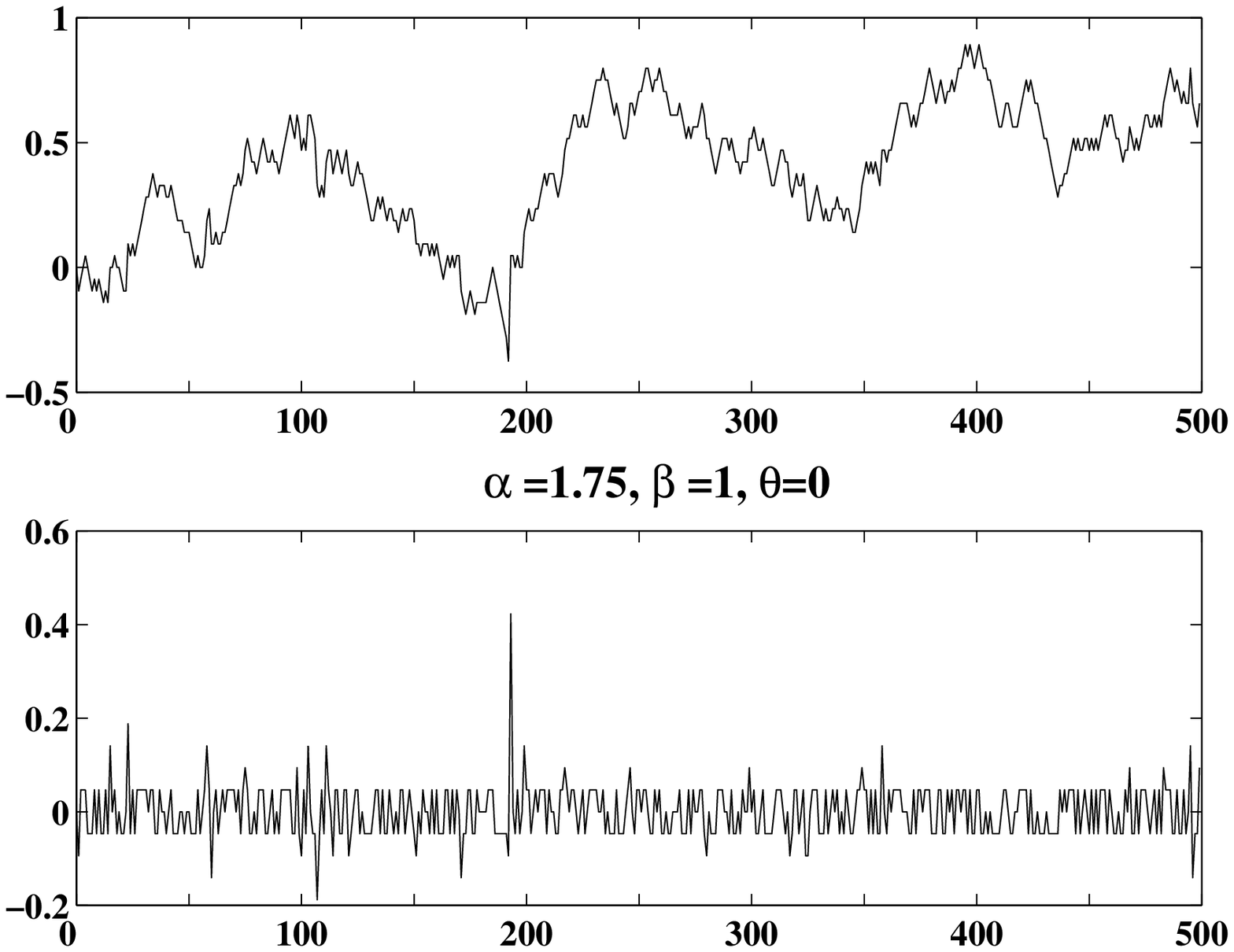}
\end{center}
\vspace{-0.2truecm}
\begin{center}
 \includegraphics[width=.48\textwidth]{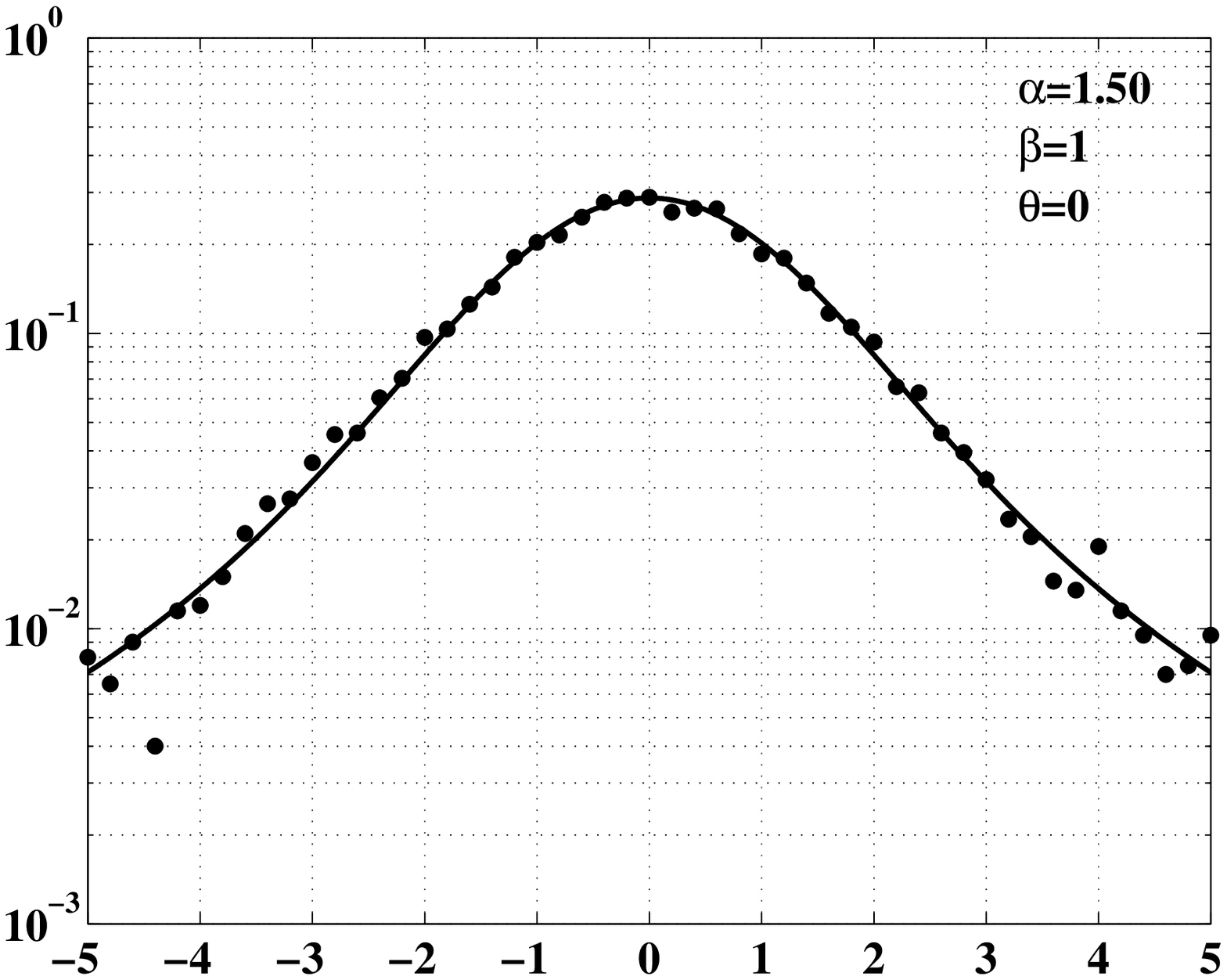}
 \includegraphics[width=.48\textwidth]{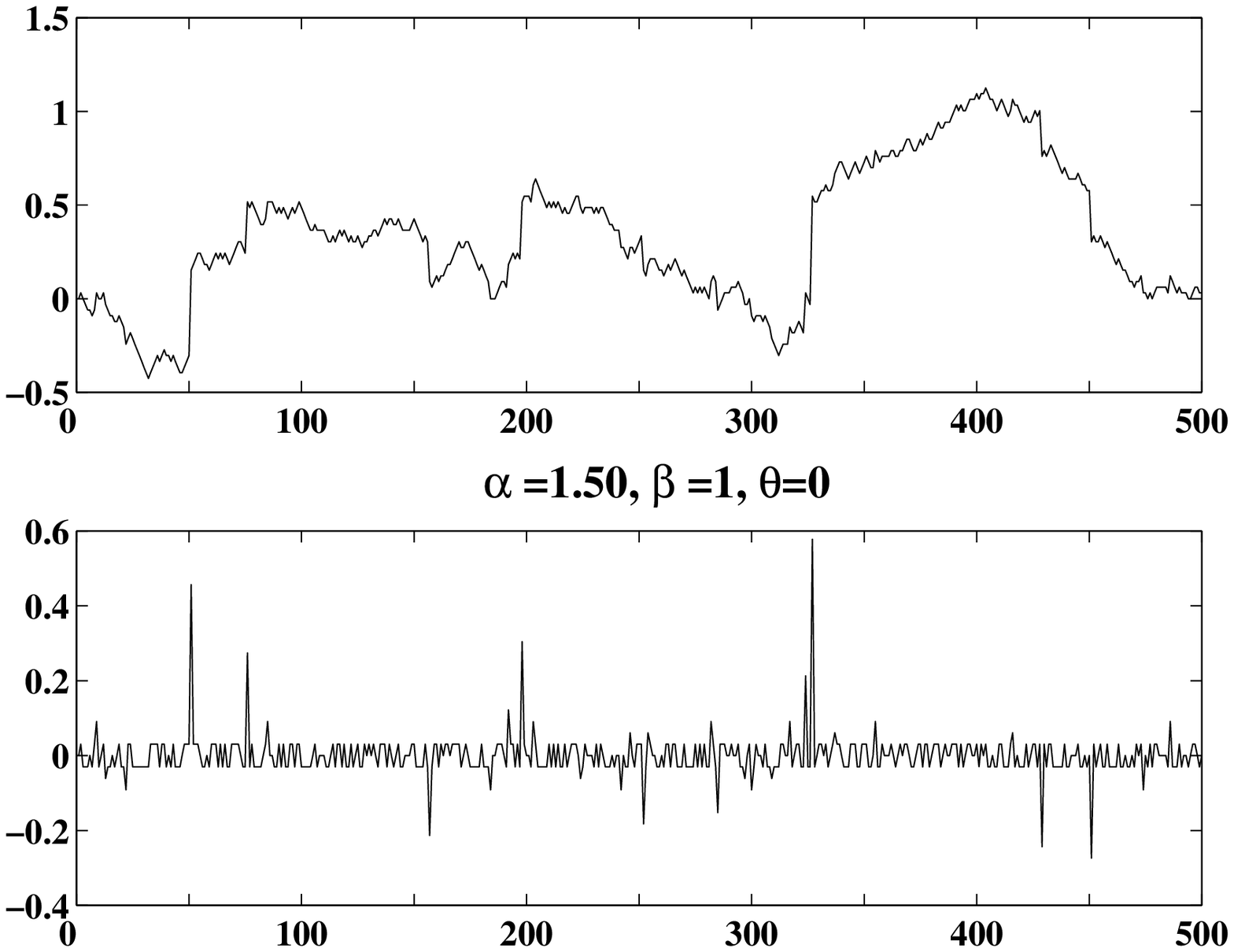}
 \end{center}
 \vspace{-0.2truecm}
\caption{Histograms (left) and sample paths with increments (right) for symmetrical, strictly space fractional diffusion.}
\end{figure}
 \begin{figure}[!ht]
 \begin{center}
 \includegraphics[width=.48\textwidth]{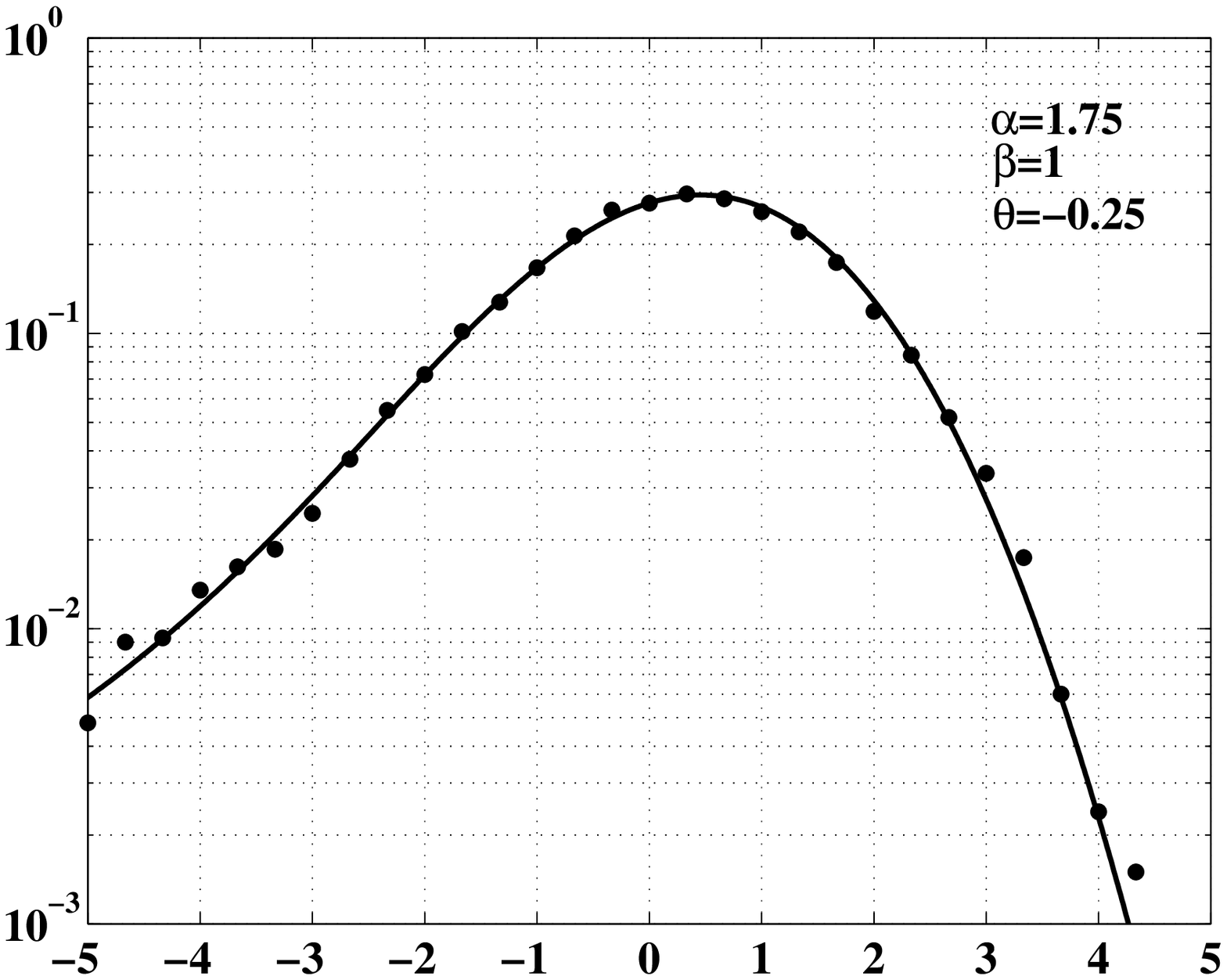}
 \includegraphics[width=.48\textwidth]{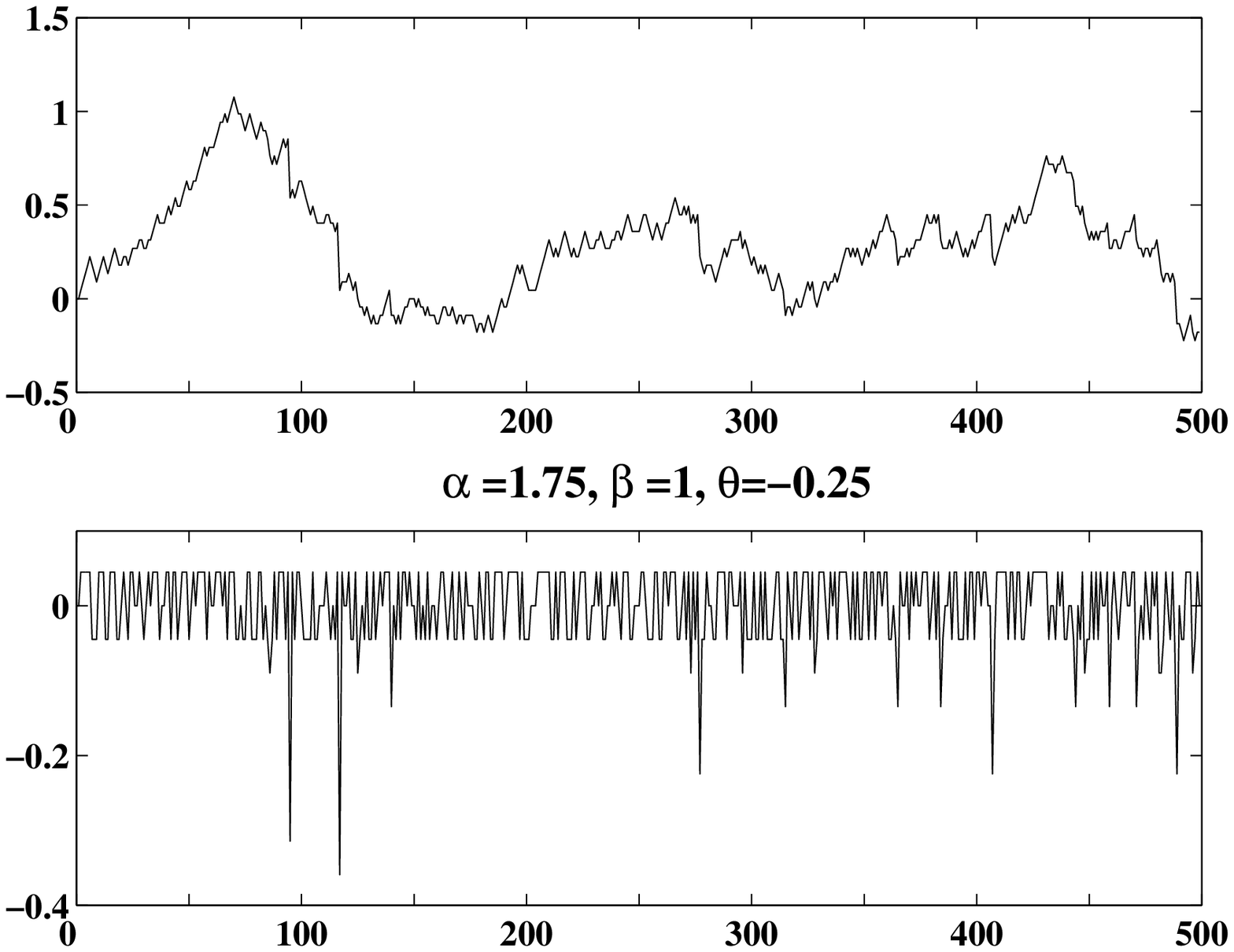}
 \end{center}
\vspace{-0.2truecm}
\begin{center}
 \includegraphics[width=.48\textwidth]{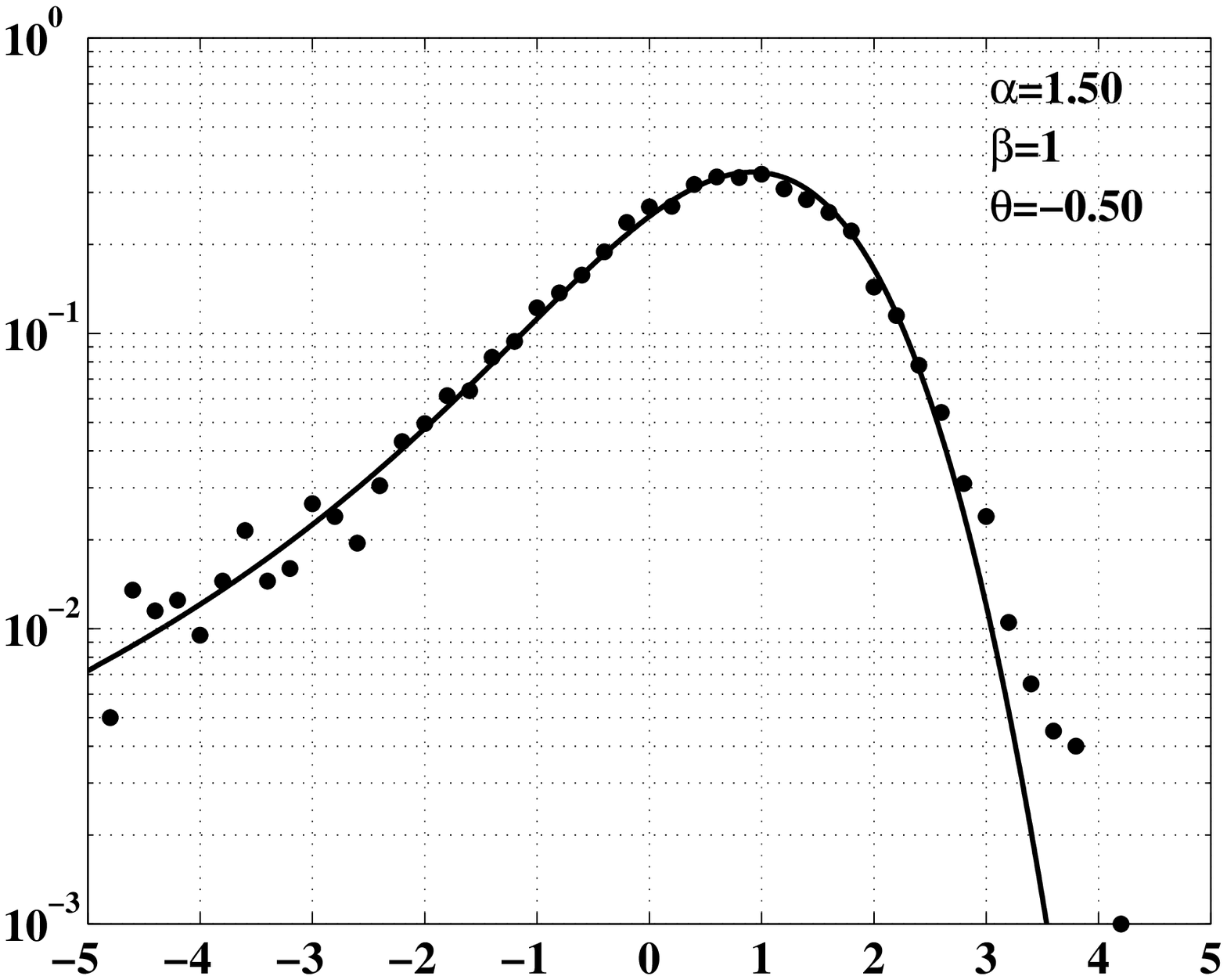}
 \includegraphics[width=.48\textwidth]{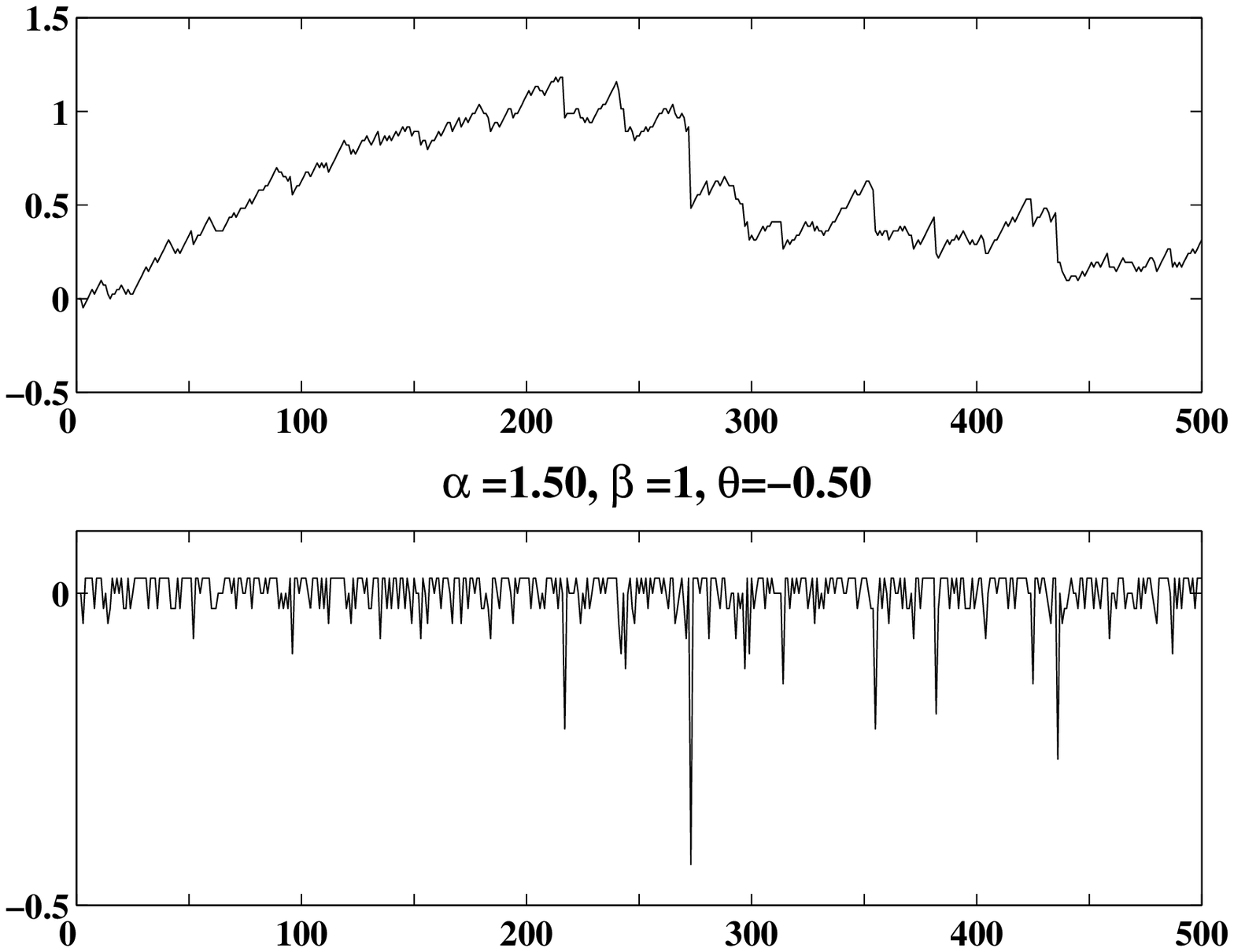}
\end{center}
\vspace{-0.2truecm}
\caption{Histograms (left) and sample paths with increments (right) for extremal, strictly space fractional diffusion.}
\end{figure}
\vsp
The plates in Fig. 7 are concerning  two
cases of {\it strictly time fractional diffusion}:
$\{\alpha =2,\, \beta  = 0.75\}$ and $\{\alpha=2, \,\beta =  0.50\}$;
here the paths exhibit the
memory effect visible in a kind of stickiness combined with
occasional jumps to points previously occupied, in distinct contrast
to the rather tame behaviour in case $\{\alpha =2,\, \beta =1\}$
(simulation of Brownian motion).

\begin{figure}[!ht]
\begin{center}
 \includegraphics[width=.48\textwidth]{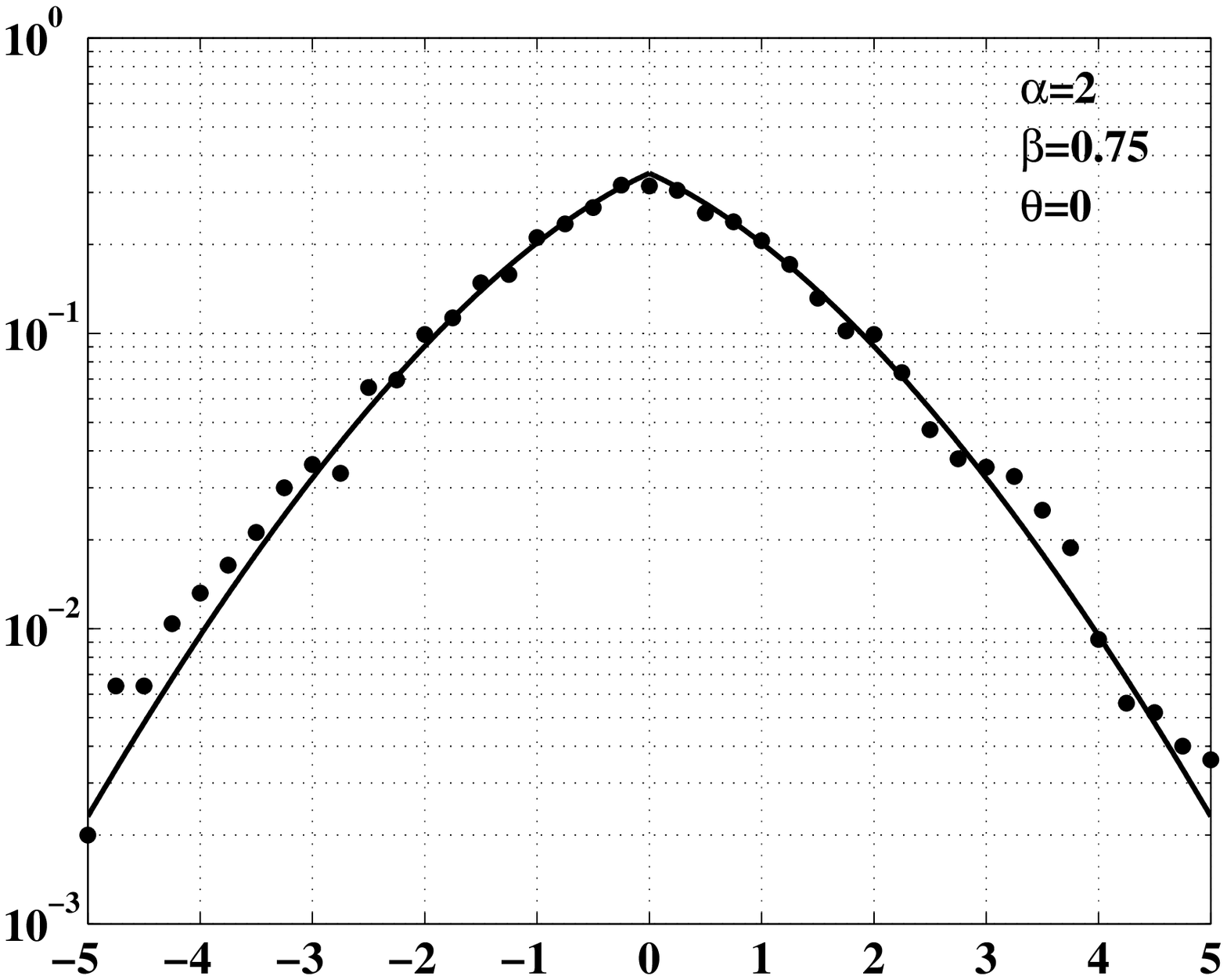}
 \includegraphics[width=.48\textwidth]{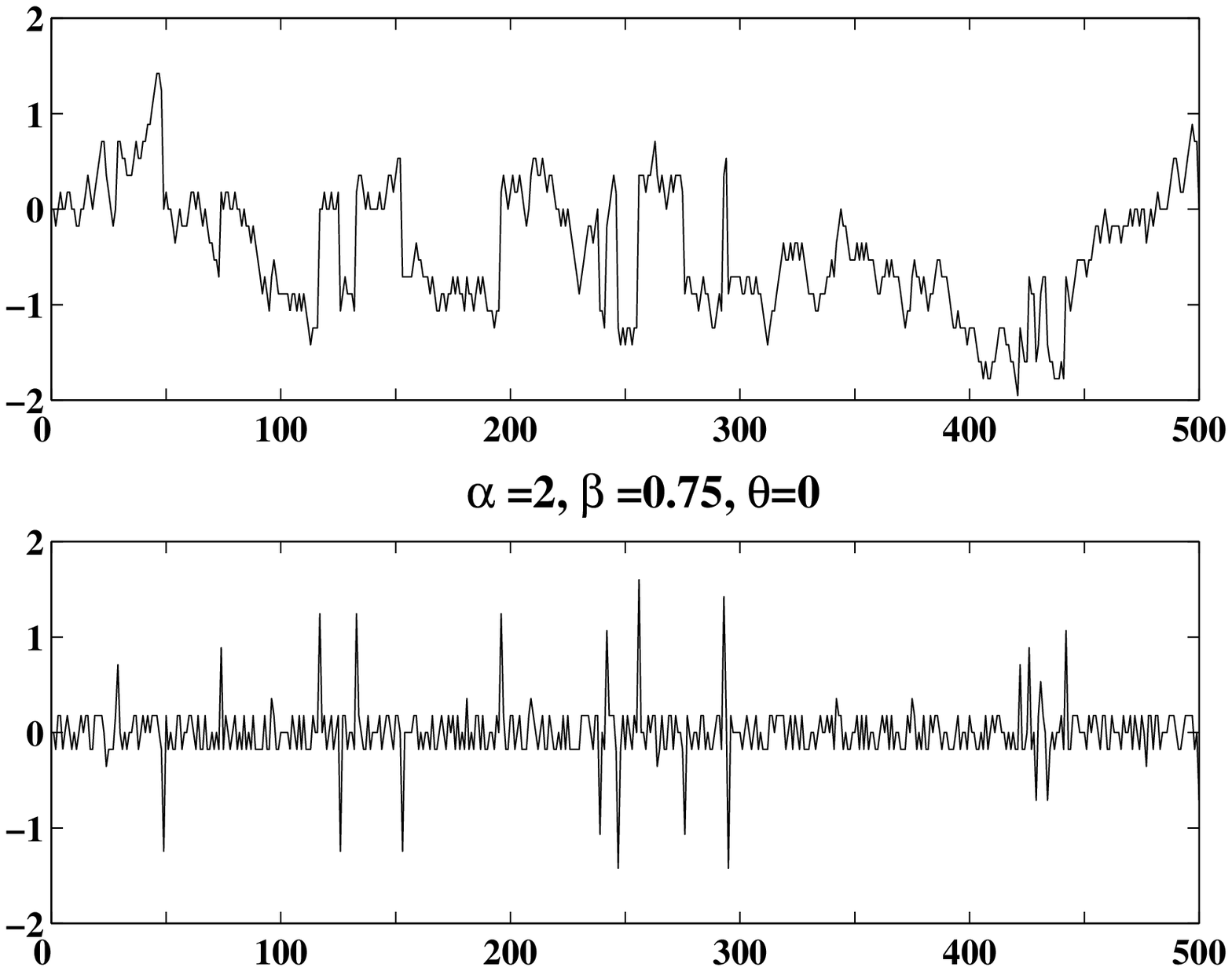}
\end{center}
\vspace{-0.2truecm}
\begin{center}
 \includegraphics[width=.48\textwidth]{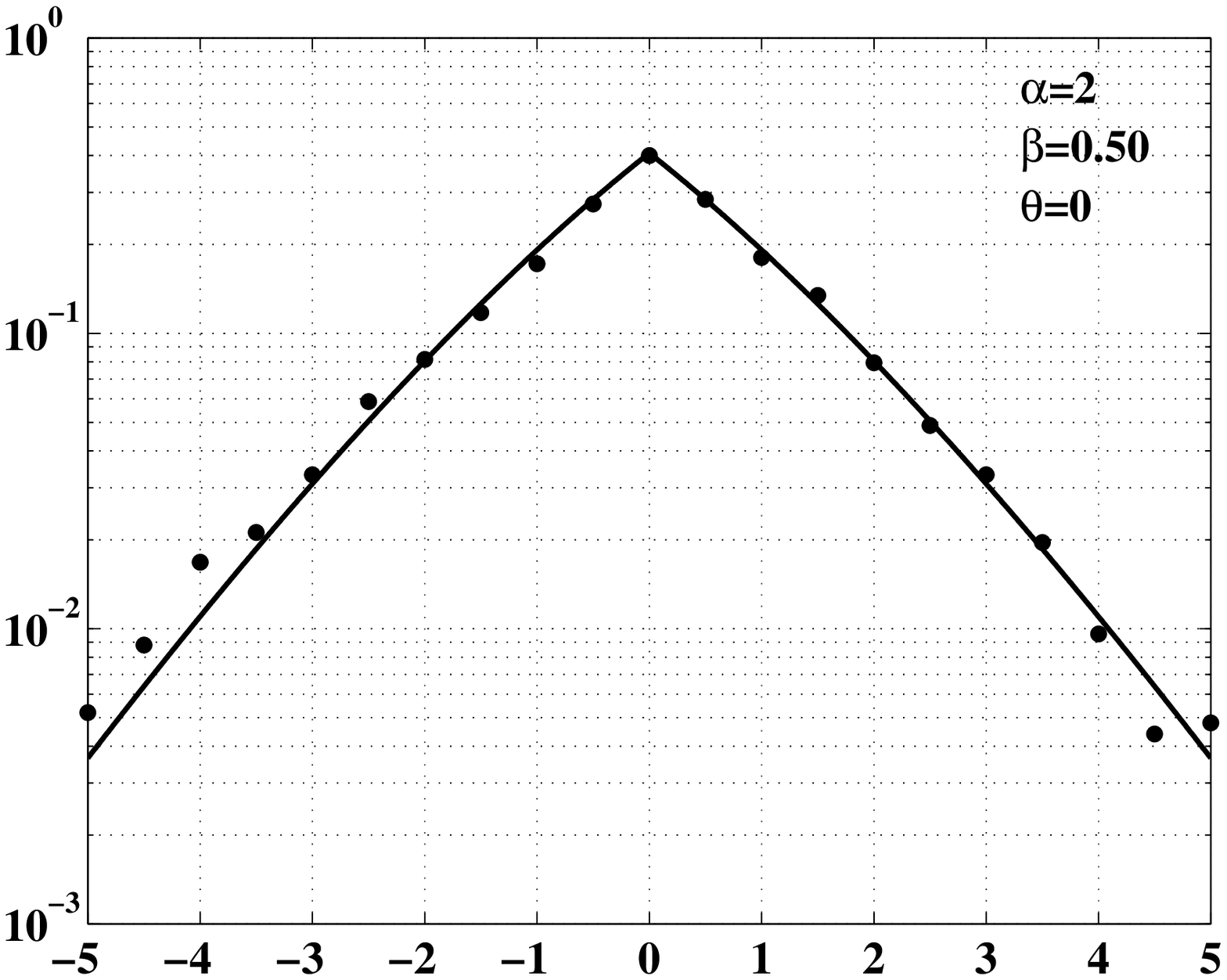}
 \includegraphics[width=.48\textwidth]{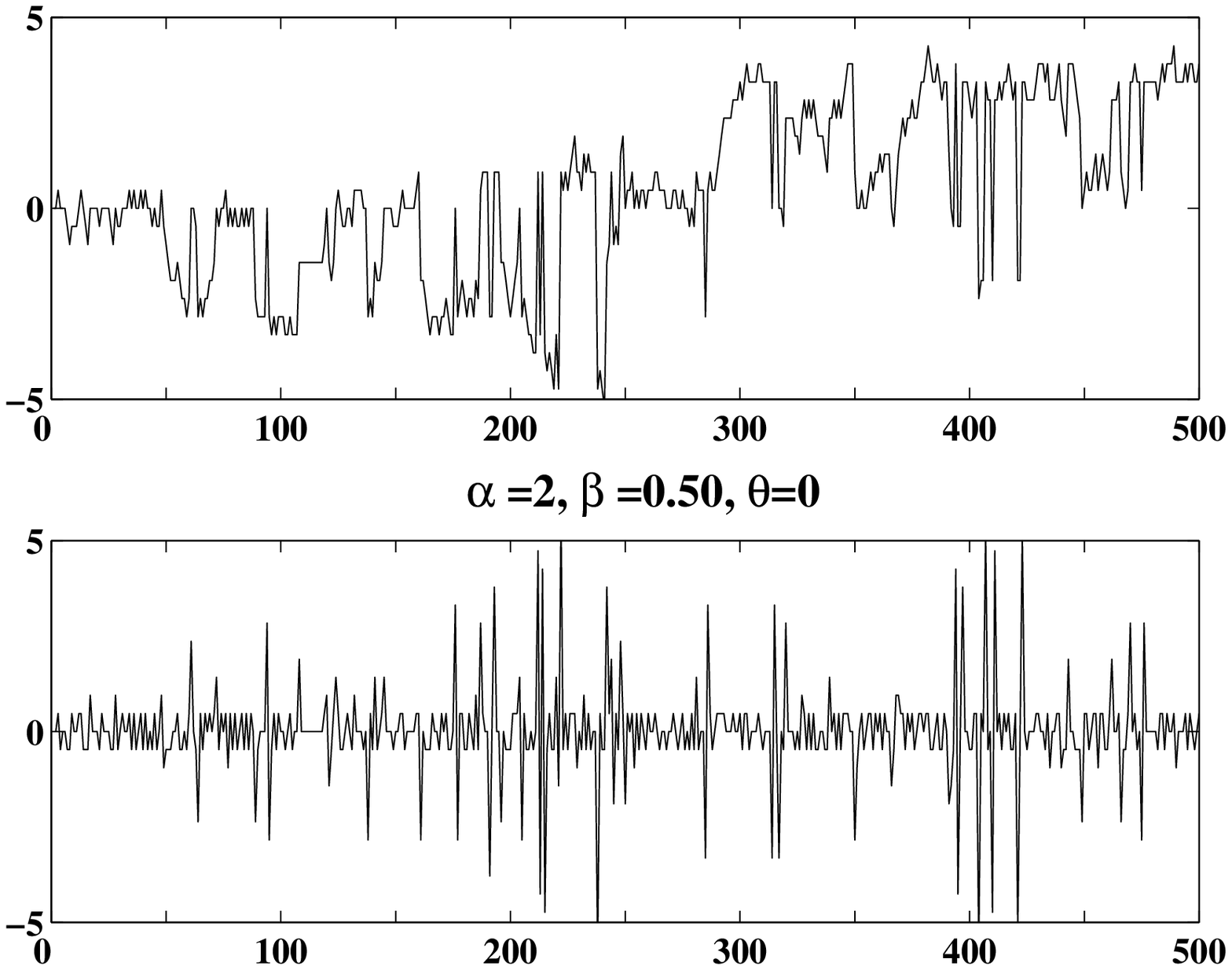}
\end{center}
\vspace{-0.2truecm}
\caption{Histograms (left) and  sample paths with increments (right)
for strictly time fractional diffusion.}
\end{figure}
\vsp
Finally, the plates in Fig. 8 are concerning
two cases of {\it strictly space-time fractional diffusion}:
$\{\alpha= 1.50,\,\beta =0.50, \, \theta=0\}$ and
$\{\alpha= 1.50,\,\beta =0.50, \, \theta=-0.50\}$,
where the combined effects of the previous cases are present.
 \vsp
 We have used our discrete models for simulation of particle
trajectories by interpreting our redistribution schemes as
descriptions of Markov	chains with
 infinitely many states,
namely the possible positions $x_j$. However, as they have the form of
difference schemes in a regular space-time grid they can "in principle"
also serve for the purpose of approximate computation of the temporal
evolution of the density $u(x,t)$. Namely, we are expecting that
$u(x_j,t_n)$ is approximated by $y_j(t_n)/h$. In fact, from their
essential properties (conservativity and preservation of
non-negativity, see (4.6), (5.11) and (5.12) ) it can by standard
methods of numerical analysis been shown that our models interpreted
as difference schemes are stable and consistent, hence convergent.
We say "in principle" because for practical application appropriate
truncations are required.
That there is also convergence in the chain interpretation
(complete convergence in the stochastic sense) can be shown by
methods of Fourier and Laplace analysis, see Gorenflo and Mainardi
in \cite{GorMai ZAA99,GorMai CHEMNITZ01}
for the space fractional diffusion. 
\begin{figure}[!ht]
\begin{center}
 \includegraphics[width=.48\textwidth]{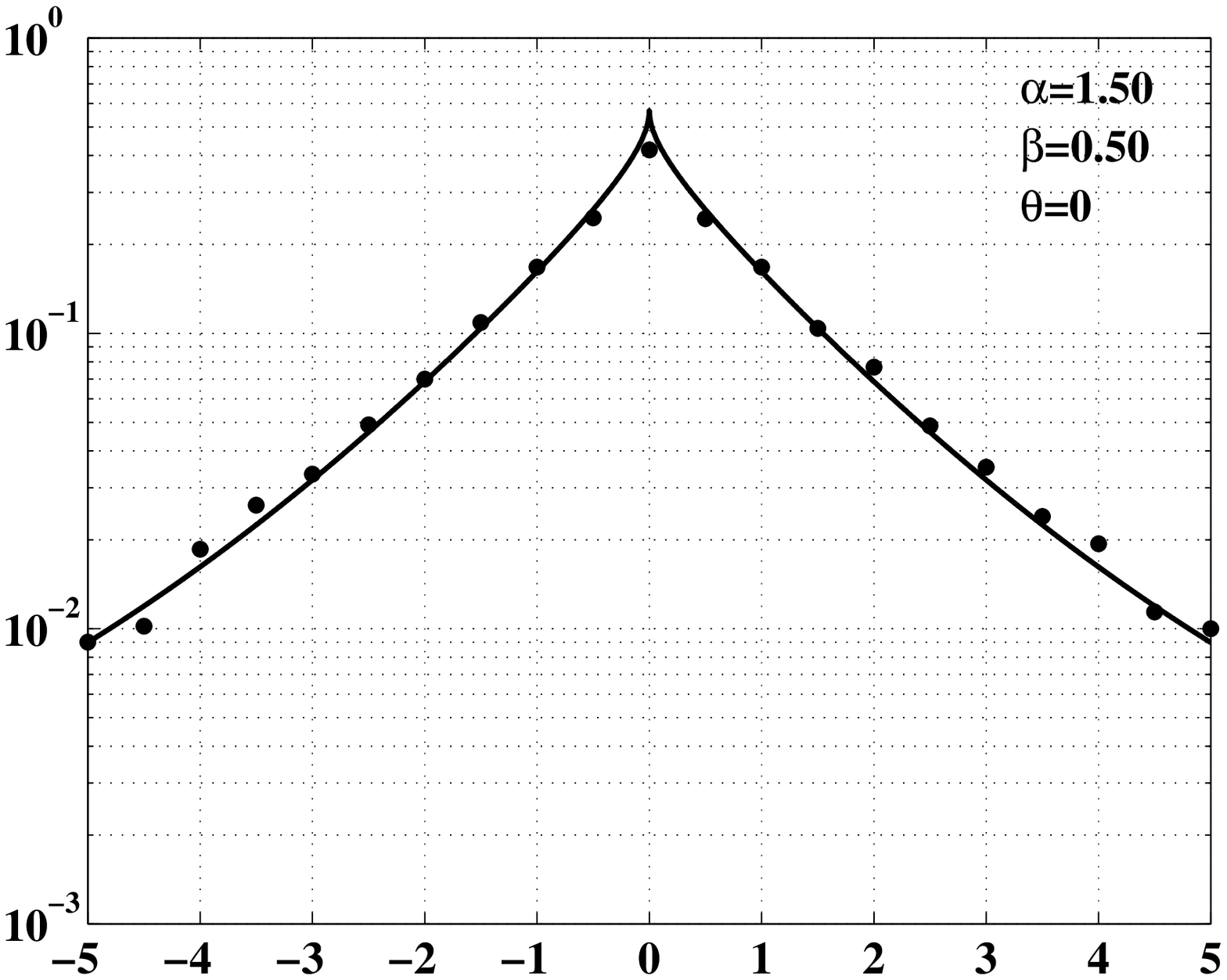}
 \includegraphics[width=.48\textwidth]{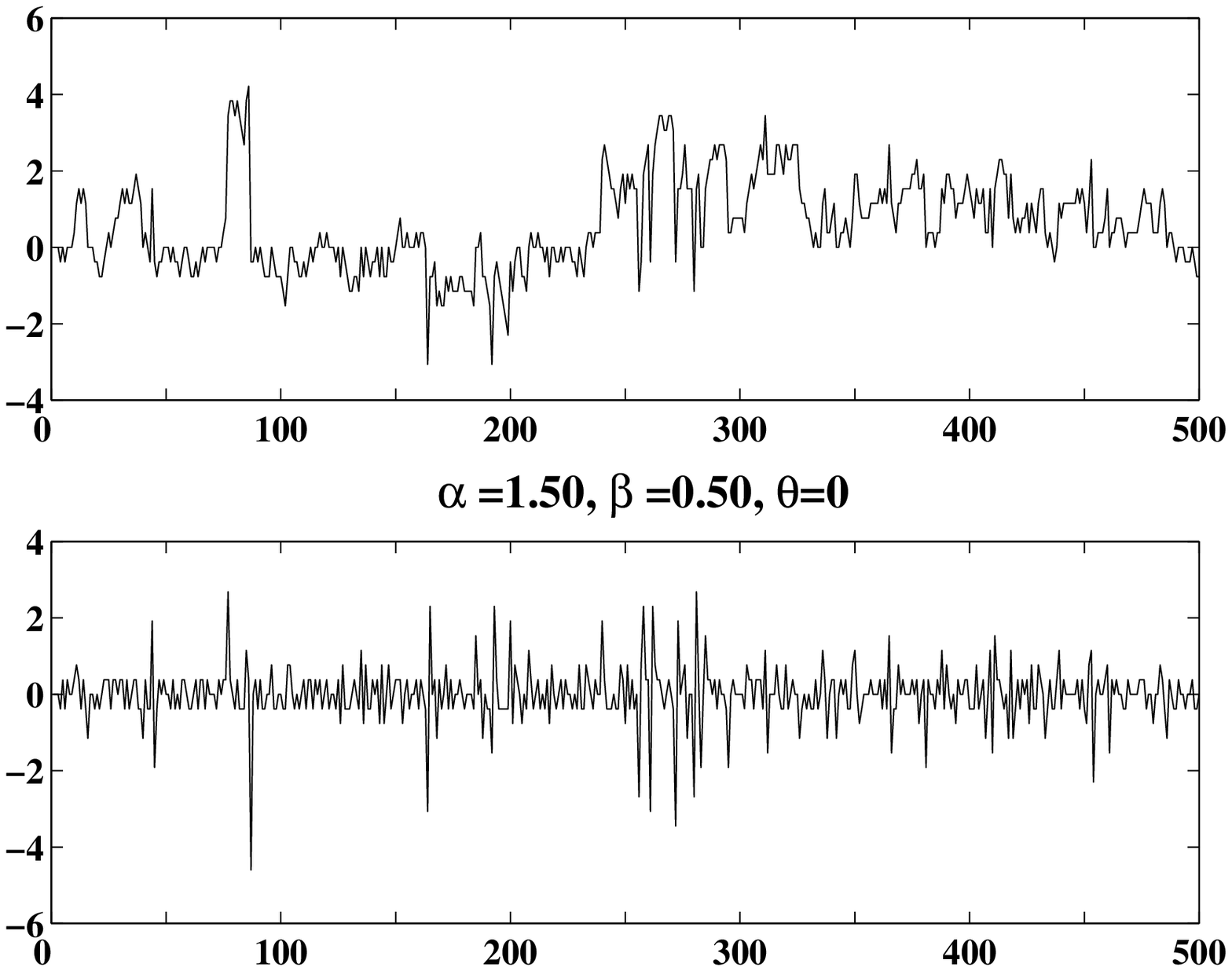}
\end{center}
\vspace{-0.2truecm}
\begin{center}
 \includegraphics[width=.48\textwidth]{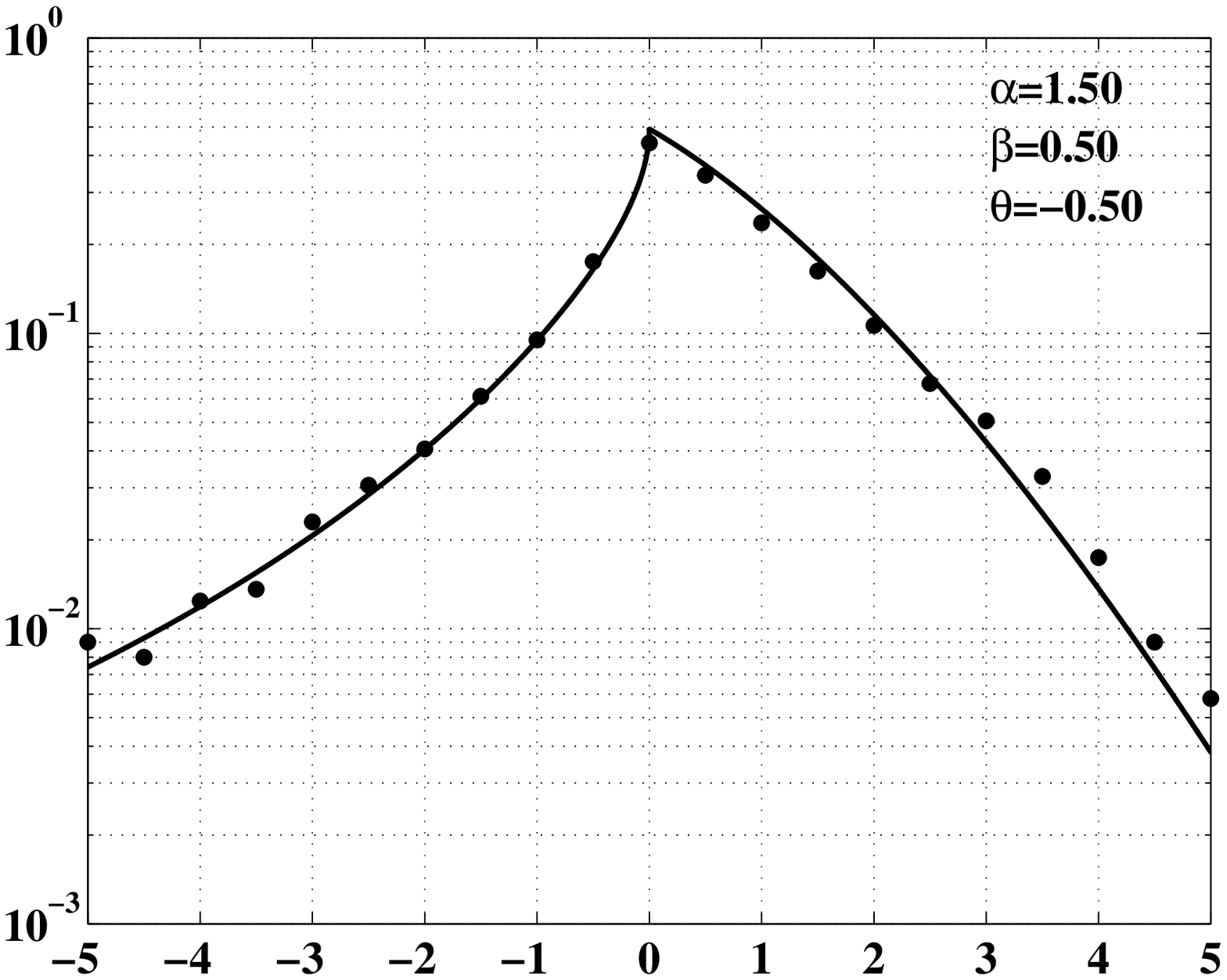}
 \includegraphics[width=.48\textwidth]{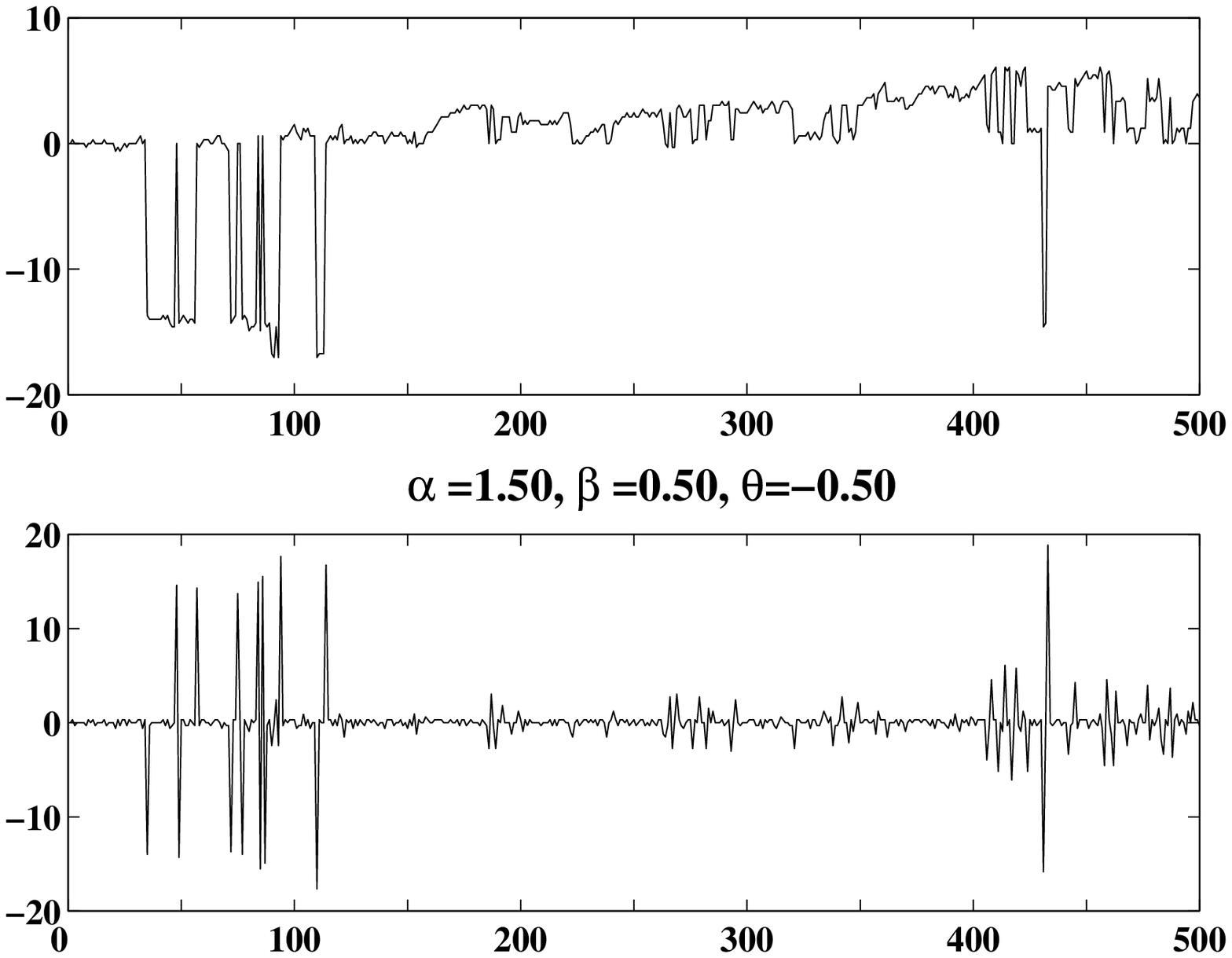}
\end{center}
\vspace{-0.2truecm}
\caption{Histograms (left) and  sample paths with increments (right)
for strictly space-time fractional diffusion.}
 \end{figure}
\vsp
There are other methods of simulating space-fractional diffusion
processes. Without attempting to be  exhaustive, without going into
details and without attempting to give a survey of the existing and
growing literature on the subject let us mention a few possibilities:
generations of random numbers distributed according to a given stable
law, random walks discrete in time but continuous in space (namely
jumps to arbitrary places in space, not only to grid points).
Furthermore, let us
mention here the Chechkin-Gonchar random walk (suggested and used in
\cite{ChechkinGonchar PhysA00}
and mathematically analyzed in \cite{GorMai CHEMNITZ01}),
and simulation via compound Poisson
processes (with proper scaling of space and time),
see \cite{GorMai POISSON02}.
 \vsp
In the special
field of methods, discrete
in space and time, there also are other methods available
(see the methods sketched in \cite{GorDFMai PhysA99}
and discussed in detail in \cite{GorMai ZAA99,GorMai CHEMNITZ01}
of which, in particular, the Gillis-Weiss random walk must be cited,
see \cite{GillisWeiss 70}).
Final word: Still many questions are open in this challenging and
fascinating field of research!

\newpage
\noindent
\subsection*{Appendix A: The Riesz-Feller  space fractional  derivatives}

In this Appendix we provide the explicit expression of the 
{\it Riesz-Feller} fractional derivative $\, _xD^\alpha_\theta$
which, according to (2.4)  is defined as
the  pseudo-differential operator with symbol
$$\widehat{\, _xD^\alpha_\theta}
  = {\ds -|\kappa|^\alpha} \, \e^{\ds  i (\sgn \kappa)\theta\pi/2}\,.
\eqno(A.1)$$
Let us now express more properly the operator 
$\, - \,_xD_\theta ^\alpha $
 as  inverse
of a suitable integral operator $\,_xI_\theta ^\alpha \,, $
 whose symbol is
required to be
$\,{\ds |\kappa |^{-\alpha}}\, 
\e^{\ds \,-i\,({\sign}\,\kappa)\,\theta \pi/2}\,,$
so we may write
$$ _xD_\theta^\alpha := - \, _xI_\theta ^{-\alpha}\,. \eqno(A.2)$$
This integral operator was  found by Feller 
\cite{Feller 52} in 1952  generalizing the approach by
Marcel Riesz to Fractional Calculus, see \eg \cite{Riesz 49},
and it is referred to as
{\it Feller potential}
by  {Samko, Kilbas \&  Marichev} \cite{SKM 93}
\footnote{
We must note that in his original paper Feller 
used a  skewness parameter $\delta $ different from our $\theta \,; $
the symbol of the Feller potential   is
$$  \l(|\kappa|\,\e^{\ds\,-i\,({\sign}\,\kappa)\,\delta}\r)^{-\alpha}
\,,  \q \hbox{{\rm so}}
 \q \delta =  -{\pi\over 2}\, {\theta \over\alpha}\,,
\q \theta = - {2\over \pi} \, \alpha \delta \,. $$
  Feller and Samko, Kilbas \& Marichev thus use $I^\alpha _\delta $
where their $\delta $ is related to our $\theta $ as above.}.
Using our notation, the {\it Feller potential} reads
$$\,_xI_\theta ^\alpha \, f(x) =
  c_-(\alpha,\theta) \,_xI_+^\alpha \,f(x)
+ c_+(\alpha,\theta) \,_xI_-^\alpha \, f(x) \,,
   \eqno(A.3)$$
where, if $0<\alpha <2\,,\; \alpha \ne 1\,, $
$$ 
c_+(\alpha,\theta) =
  {\sin \,\l[(\alpha-\theta )\, \pi/2\r] \over\sin\,(\alpha\pi)}\,,
 \q
   c_-(\alpha,\theta) =
{\sin\,\l[(\alpha+\theta )\, \pi/2\r] \over \sin(\alpha\pi)} \,,
 \eqno(A.4)$$
and, by passing to the limit (with $\theta =0$)
$ c_+(2,0) = c_-(2,0) = - 1/2\,. $
In (A.4)  the operators $\,_xI_\pm^\alpha $ denote
the {\it Riemann-Liouville fractional integrals}, also known as
{\it Weyl fractional integrals}), defined as
$$\cases{
 {\ds \,_xI_+^\alpha \, f(x)} =
{\ds \rec{\Gamma(\alpha)}}\,
   {\ds \int_{-\infty}^x \!\! (x-\xi)^{\alpha-1}\,f(\xi)\,d\xi} \,,
 \cr\cr
{\ds \, _xI_-^\alpha \, f(x)} =
{\ds \rec{\Gamma(\alpha)}}\,
 {\ds \int_x^{+\infty} \!\! (\xi-x)^{\alpha-1}\,f(\xi)\,d\xi} \,.\cr}
  \eqno(A.5)$$
We note that whereas the coefficients $c_\pm$ can loose their meaning
when $\alpha$ is an integer, the Riemann-Liouville integral operators
$I_\pm^\alpha$
are well defined in their action on rapidly decreasing
integrable functions, for any  $\alpha \ge 0\,, $
being set equal to the identity operator  when $\alpha =0\,,$
for convenience (justified by passage to the limit $\alpha \to 0$).
In the particular case $\theta =0$ we get
$$c_+(\alpha ,0)  = c_-(\alpha ,0) =
  {1 \over 2\, \cos\,(\pi\alpha /2)} \,,\eqno(A.6)$$
and
thus we  recover the {\it Riesz potential},
see \eg \cite{SKM 93},
$$ \,_xI_0^\alpha \, f(x) :=
\rec{2\, \Gamma(\alpha )\, \cos\,(\pi\alpha /2)}\,
\int_{-\infty}^{+\infty}
     \!\!\! |x-\xi |^{\alpha -1}\,f(\xi )\, d\xi
 \,.  \eqno(A.7)$$
The Riesz and the Feller potentials are well defined if the
index is located in the interval  $(0\,,\, 1)\,, $ 
or in the interval $(0,2)\,, $ and we have  the
semigroup property,
$\, _xI_\theta^\alpha \,\, _xI_\theta^\beta
= \, _xI_\theta^{\alpha+\beta}  \,,$ if
$0<\alpha<1$,  $0<\beta<1$ and  $\alpha+\beta <1 \,. $
Then, following Feller, we   define
by analytic continuation the pseudo-differential operator   
(A.2) in the whole range $0<\alpha \le 2\,,\, \alpha \ne 1\,,$  as
$$
 \,_xD_\theta^{\alpha}  :=
 -\l[ c_+(\alpha,\theta)\,\,_xD_+^{\alpha} +
c_-(\alpha,\theta)\,\,_xD_-^{\alpha}\r]\,,\q 0<\alpha \le 2 \,.
  \eqno(A.8)
$$
where
 $  \,_xD_+^{\alpha}:=\,_xI_+^{-\alpha}$ and
 $  \,_xD_-^{\alpha}:=  \,_xI_-^{-\alpha}$
are the  inverses of the operators
$\,_xI_+^\alpha$ and  $\,_xI_-^\alpha \,,$ respectively,
and are referred to as the {\it Weyl fractional derivatives}.
When $\theta =0$  the {\it Riesz-Feller derivative} 
can be simply referred to as    {\it Riesz derivative}. 
We note from (A.4) the property
$ \, c_\pm(\alpha, \theta) = c_\mp (-\alpha)\,. $
For integral representations of the  operators
$\,_xI_\pm^{-\alpha}$ 
see \cite{SKM 93}; we have
$$
_xI_\pm^{-\alpha} = \cases{
     {\ds \pm {d \over dx}} \,\l(\,_xI_\pm^{1-\alpha}\r) \,,
   & if $\q 0<\alpha \le 1 \,,$\cr\cr
     {\ds{d^2 \over dx^2}} \,\l(\,_xI_\pm^{2-\alpha}\r) \,,
   & if $\q 1<\alpha \le 2 \,.$\cr}
\eqno(A.9)
$$
For $\alpha =2$ ($\theta =0 $) we recover the standard second derivative.
In fact, from (A.6)
$c_+ (2,0) = c_- (2,0)= -1/2\,,$
so from (A.8)-(A.9)
$ _xD_0^2 = - I_0^{-2}
   = \l(\,_xI_+^{-2} +  \,_xI_-^{-2}\r)/2 
   =   \l( {d^2 \over dx^2} + {d^2 \over dx^2}\r)/2 = {d^2 \over dx^2}
=\, _xD^2 \,. $
For $\alpha =1$ ($|\theta| \le 1$) we have
(from calculation with symbols) 
$$ _xD^1_\theta \, f(x) = \l[ \cos (\theta \pi/2)\,  _xD_0^1
 + \sin (\theta \pi/2)\,  _xD \r] \, f(x) \,,\eqno(A.10)$$
where
$ \, _xD_0^1 \, f(x) = -\,_xD \,\l[\, _xH \, f(x)\r] \,,$
with 
$$ \,_xH \, f(x) =  {1\over \pi}
\,  \int_{-\infty}^{+\infty} {f(\xi )\over x-\xi}\, d\xi
=    {1\over \pi}
\,  \int_{-\infty}^{+\infty} {f(x-\xi )\over \xi}\, d\xi
 \,.
\eqno(A.11) $$ 
In (A.11)  $\,_xH$ denotes the Hilbert transform
(with symbol $\widehat{\,_xH}(\kappa) = i\, \sgn\, \kappa\,$) 
and its singular integral is understood in the Cauchy principal value
sense.
We note that in the limiting extremal cases $\theta = \pm 1$ we 
recover the standard first derivative, \ie
$\,_xD^1_{\pm 1} = \pm\, \, _xD\,.$ 
\vfill\eject
\subsection*{Appendix B: The Riemann-Liouville and Caputo time fractional
derivatives}
\vsp
The purpose of this Appendix is to clarify for the interested reader
the main differences between
the {\it Caputo} fractional derivative
$\, _tD_*^\beta$ adopted in this paper, see (2.8),
and the {\it Riemann-Liouville} fractional derivative
$\, _tD^\beta$	usually adopted in the literature.
Any formula here is valid for $t>0\,, $
with the assumption that the function $f(t)$ has
a finite limit as $t \to 0^+\,. $
We recall
$$
 _tD^\beta  \,f(t) := \cases{
  {\ds	\rec{\Gamma(1-\beta)}}\,
  {\ds {d \over dt}}\,
 {\ds \int_0^t
    {f(\tau)  \over (t-\tau )^\beta}\,d\tau} \,,
  & $\; 0<\beta  <1\,,$\cr\cr
 {\ds {d \, f(t) \over dt}}\,, & $\; \beta =1\,.$\cr}
\eqno(B.1) $$
The two  fractional derivatives  are related to
the Riemann-Liouville fractional  integral as follows.
The Riemann-Liouville fractional  integral    is
$$ _tJ^\mu \, f(t) :=
\rec{\Gamma(\mu )}\,\int_0^t
    {f(\tau)  \, (t-\tau )^{\mu -1}}\,d\tau \,,\q \mu >0\,,
  \eqno(B.2) $$
(with the convention $\,_tJ^0 \, f(t) = f(t)$)
and is known to satisfy the semigroup property
$ _tJ^\mu\, _tJ^\nu = \,_tJ^{\mu+ \nu }\,, $
with $\mu, \nu >0\,.\,$
For any $\beta >0$
the Riemann-Liouville fractional derivative is
defined as the {\it left} inverse of the corresponding
fractional integral
(like  the derivative of any integer order), namely
$   \, _tD^\beta \, _tJ^{\beta}\, f(t)= f(t)\,.$
Then for  $\beta \in (0,1] $ we have
 $$   _tD^\beta\, f(t) := _tD^1\, _tJ^{1-\beta}\, f(t)\,,\q
      _tD_*^\beta f(t) := _tJ^{1-\beta} \, _tD^1\, f(t)\,,
\eqno(B.3)$$
$$
_tJ^\beta \, _tD_*^{\beta}\, f(t)=
 \,_tJ^\beta\, _tJ^{1-\beta}\, _tD^1\, f(t) =
  \,_tJ^1\, _tD^1\, f(t) = f(t) - f(0^+)\,.
\eqno(B.4)
$$
Recalling the rule
$$ _t D^{\beta }\, t^{\gamma}=
   {\Gamma(\gamma +1)\over\Gamma(\gamma +1-\beta )}\,
     t^{\gamma-\beta }\,,
 \q \beta  \ge 0\,,
  \q \gamma >-1\,, \eqno(B.5)
$$
it turns out  for  $0 <\beta \le 1\,,$
$$ _tD_*^\beta	\,f(t)	\, = \, _tD^\beta  \,\l[ f(t) -
   f(0^+) \r] =
\, _tD^\beta  \, f(t) -
      f(0^+) \,
{t^{-\beta }\over \Gamma(1-\beta)}\,.
 \eqno(B.6) $$
Note that for $\beta=1$ the two types of fractional derivative coincide.
\vsp
The {\it Caputo} fractional derivative
represents a sort of regularization in the time origin for the
{\it Riemann-Liouville} fractional derivative
and  satisfies the  relevant property
of being zero when applied to a constant.
For more details on this fractional derivative (and its
extension to higher orders) we refer the interested reader to
Gorenflo and Mainardi \cite{GorMai CISM97},
Podlubny \cite{Podlubny 99} and
Butzer and  Westphal\cite{ButzerWestphal 00}.

\vfill\eject

\subsection*{Acknowledgements}

We are grateful to the Italian Group of Mathematical Physics (INDAM),
to the Erasmus-Socrates project
and to the Research Commissions of the
Free University of Berlin ({\it Convolutions} Project)
and of the University of Bologna
for supporting joint efforts of
our research groups in Berlin and Bologna.
We also acknowledge partial support by the INTAS Project 00-0847.


\end{document}